\documentclass[epj]{svjour}
\usepackage{latexsym}
\usepackage{graphicx}
\usepackage{color}
\usepackage{amssymb}
\usepackage{amsmath}
\usepackage{hyperref}
\usepackage{float}
\usepackage{url}
\usepackage{subfig}

\usepackage{longtable}

\usepackage{lineno, xcolor}
\begin{document}

\title{Neutral-Current Weak Interactions at an EIC}

\author{
	Y. X. Zhao\inst{1}\thanks{\emph E-mail: yuxiang.zhao@stonybrook.edu}, 
	A. Deshpande\inst{1}, 
	J. Huang\inst{2}, 
	K. S. Kumar\inst{1}, 
	S. Riordan\inst{1}
}            

%
%

\institute{Department of Physics and Astronomy, Stony Brook University, Stony Brook, NY 11794, USA 
\and Physics Department, Brookhaven National Lab, Upton, NY 11793, USA }
\date{\today}


\abstract{
A simulation study of measurements of neutral current structure functions of the nucleon at the future high-energy and high-luminosity 
polarized electron-ion collider (EIC) is presented.  A new series of $\gamma-Z$ interference structure 
functions, $F_1^{\gamma Z}$, $F_3^{\gamma Z}$, $g_1^{\gamma Z}$, $g_5^{\gamma Z}$ become accessible via 
parity-violating asymmetries in polarized electron-nucleon deep inelastic scattering (DIS).
Within the context of the quark-parton model, they 
provide a unique and, in some cases, yet-unmeasured combination of unpolarized and polarized 
parton distribution functions.  The uncertainty projections for these 
structure functions using electron-proton collisions are considered for various EIC
beam energy configurations.  Also presented are uncertainty projections for measurements of
the weak mixing angle $\sin^2 \theta_W$  using electron-deuteron collisions which cover
a much higher $Q^2$ than that is accessible in fixed target measurements.  QED and QCD 
radiative corrections and effects of detector smearing are included with the calculations. 
}


\authorrunning{Y. X. Zhao {\it et al.}}
\titlerunning{Neutral Weak Interactions at an EIC}
\maketitle

\section{Introduction}
The proposed polarized electron-light ion collider~\cite{Accardi:2012qut} offers unprecedented new 
opportunities to study the QCD structure of the nucleon due to the combination of the high luminosity,
high center of mass energy, and polarized electron and hadron beams, that it is expected to provide. 
Extensive studies of polarized electron-nucleon 
deep inelastic scattering at such a new facility have been carried out, accessing spin structure functions of 
the nucleon at low Bjorken-$x$ via longitudinal double-spin asymmetries~\cite{Aschenauer:2012ve,DeRoeck:1997gy,Kuhn:2008sy}, and  weak
charged current interactions~\cite{Aschenauer:2013iia}.

In this paper, such studies are extended to include new parity-violating single-spin observables.
At the proposed EIC where momentum transfers $Q^2$ is large but still $ \lesssim M_Z^2$, parity-violating 
asymmetries will be dominated by weak-electromagnetic (so-called $\gamma-Z$) interference amplitudes 
that could provide new and complementary information about the structure of the nucleon. New linear combinations of 
$u$-, $d$- and $s$-quark/anti-quark distribution functions would become accessible, potentially enabling a 6-flavor separation of parton distribution functions in an interesting kinematic region.

This paper presents the formalism and defines the observables of interest in Sec~\ref{sec:formalism}, then describes
Monte Carlo simulation studies in Sec.~\ref{simulation}, and finally in Sec.~\ref{sec:results} the subsequent extraction 
of the uncertainty projections for the new neutral current structure functions and weak mixing angle over the
range of relevant kinematics. 

\section{Neutral Currents in the DIS region}
\label{sec:formalism}
In electron-hadron interactions, the differential scattering distribution consists of contributions
from both virtual photon and Z boson exchanges as well as their interference: $d\sigma \sim |M_{\gamma} + M_{Z}|^2$. 
One can write down the generic lepton-hadron interaction tensor in terms of structure functions in the 
DIS kinematic region under the assumption of CP-symmetry including the spins of the both, the lepton and the hadron.
For purely electromagnetic scattering at leading twist, where parity-violation is forbidden, the  deep inelastic
cross-section is related to four structure functions $F_1^{\gamma}$, $F_2^{\gamma}$, $g_1^{\gamma}$,
$g_2^{\gamma}$.  Including the weak neutral current interaction
where parity-violation is allowed, one can access the additional $\gamma-Z$ interference structure functions
$F_1^{\gamma Z}$, $F_3^{\gamma Z}$, $g_1^{\gamma Z}$ and $g_5^{\gamma Z}$ by measuring parity-violating deep inelastic scattering (PVDIS)
asymmetries.  $F_1^{\gamma Z}$ and $F_3^{\gamma Z}$ are accessible using longitudinally-polarized electrons and $g_1^{\gamma Z}$ 
and $g_5^{\gamma Z}$ using longitudinally-polarized  nucleons.  In each case the polarization of the other colliding particle is averaged over their polarization states, i.e. they are unpolarized.

In the DIS process, $\rm{e} + \rm{p}(\rm{n}) \rightarrow \rm{e} + \rm{X}$, where the nucleon gets destroyed into hadronic remnant, $\rm{X}$, it is necessary to define the kinematic variables and their relations with each other for physics discussions. With $k$, $k'$ denoting the four-momenta of the incoming and outgoing electron; $p$, the 
four-momentum of a nucleon; $Q^{2}$, the squared momentum transfer to the electron; $x$, the Bjorken momentum 
fraction of the parton; $y$,  the in-elasticity, and $W_h$, the invariant mass of the produced hadronic system 
are related with each other as follows:
\begin{eqnarray}
Q^2 &=& -q^2 = -(k-k')^2, \\
x &=& \frac{Q^2}{2p \cdot q}, \\
y &=& \frac{q \cdot p}{k \cdot p}, \\
W_h &=& \sqrt{(p+q)^2}.
\end{eqnarray}

Within the Standard Model (SM), where weak neutral current interactions are mediated by the
$Z$ boson, the R-L asymmetry for the longitudinally polarized electron
scattering off unpolarized nucleon (neglecting the pure Z-exchange amplitude)\cite{Anselmino:1993tc,Anselmino:1994gn,Wang:2014bba} can be written as 
\begin{equation}
\begin{split}
A_\mathrm{PV}^\mathrm{electron} &= \frac{\sigma^{R}-\sigma^{L}}{\sigma^{R}+\sigma^{L}}  \\
&= \frac{G_F Q^2}{2\sqrt{2} \pi \alpha} [ g_A^e \frac{F_1^{\gamma Z}}{F_1^{\gamma}} + g_V^e \frac{Y_{-}}{2Y_{+}} \frac{F_3^{\gamma Z}}{F_1^{\gamma}} ],
\end{split}
\label{eq:A_beam}
\end{equation}
where $G_F$ is the Fermi constant, $\alpha$ is the fine structure constant, 
$g_A^{e}=-0.5, g_V^{e}=-0.5 + 2\sin^2 \theta_W$ are the axial and vector couplings respectively of the electron to the $Z$ boson, and 
$Y_{-}=2y-y^2, Y_{+}=y^2-2y+2$.  
Note the Callan-Gross relations \cite{Buchmuller:1987ur} have been used to simplify the additional dependence of 
longitudinal structure functions.
The (+) - (-) asymmetry, where (+)/(-) indicates whether the spin of the nucleon is aligned parallel/anti-parallel to the electron beam direction, in inclusive
scattering of an unpolarized electron on a longitudinally polarized hadron can be written as
\begin{equation}
\begin{split}
A_\mathrm{PV}^\mathrm{hadron} &= \frac{\sigma^{(+)} - \sigma^{(-)}}{\sigma^{(+)} + \sigma^{(-)}} \\
&=\frac{G_F Q^2}{2\sqrt{2} \pi \alpha} [ g_V^e \frac{g_5^{\gamma Z}}{F_1^{\gamma}} + g_A^e \frac{Y_{-}}{Y_{+}} \frac{g_1^{\gamma Z}}{F_1^{\gamma}} ].
\end{split}
\label{eq:A_target}
\end{equation}

In the quark-parton model, the $\gamma-Z$ interference structure functions can be written as linear combinations
of the unpolarized and polarized parton distribution functions (PDFs)
\begin{eqnarray}
F_1^{\gamma Z} &=& \sum_f e_{q_f} (g_V)_{q_f} (q_f + \bar{q}_f), \\
F_3^{\gamma Z} &=& 2\sum_f e_{q_f} (g_A)_{q_f} (q_f-\bar{q}_f), \\
g_1^{\gamma Z} &=& \sum_f e_{q_f} (g_V)_{q_f}(\Delta q_f +\Delta\bar{q}_f), \\
g_5^{\gamma Z} &=& \sum_f e_{q_f} (g_A)_{q_f}(\Delta q_f - \Delta\bar{q}_f).
\end{eqnarray}
Presently in the extraction of both unpolarized and polarized PDFs, many fits rely upon data that use imprecisely known fragmentation functions for flavor identification resulting in hadronic systematic uncertainties which are difficult to rely on. The important feature of the above $\gamma-Z$ interference structure functions is that they carry different effective couplings for quarks as compared to purely electromagnetic structure functions, and therefore provide crucial 
additional handles to global fits that unfold the individual quark flavors.   

To illustrate the quark flavor sensitivity, the following discussion approximates $\sin^2 \theta_W \approx 1/4$, the interference 
structure functions become
\begin{eqnarray}
F_1^{\mathrm{proton},~\gamma Z} &\approx& \frac{1}{9} ( u+ \bar{u} + d + \bar{d}  + s + \bar{s} + c + \bar{c}),   \\
F_3^{\mathrm{proton},~\gamma Z} &=& \frac{2}{3} ( u_V + c- \bar{c}) + \frac{1}{3}( d_V + s - \bar{s}),
\end{eqnarray}
\begin{equation}
g_1^{\mathrm{proton},~\gamma Z} \approx \frac{1}{9} ( \Delta u+ \Delta \bar{u} + \Delta d + \Delta \bar{d}  + \Delta s + \Delta \bar{s} + \Delta c + \Delta \bar{c}),
\end{equation}
\begin{equation}
g_5^{\mathrm{proton},~\gamma Z} = \frac{1}{3} ( \Delta u_V + \Delta c - \Delta \bar{c}) + \frac{1}{6}(\Delta d_V + \Delta s - \Delta \bar{s}),
\end{equation}
with $u_V$ and $d_V$ representing the valence quark contributions. $F_1^{\gamma Z}$ is therefore approximately proportional to the unweighted sum of the PDF contributions. 
Likewise, $g_1^{\gamma Z}$ accesses $\Delta \Sigma \equiv \Delta u+ \Delta \bar{u} + \Delta d + \Delta \bar{d}  + \Delta s + \Delta \bar{s} + ...$, 
the quark spin contribution to the nucleon spin. The $F_3^{\gamma Z}$ and $g_5^{\gamma Z}$ present the contributions of valence quarks to the nucleon momentum and spin, respectively.

Another attractive feature of PVDIS is that it is virtually independent of hadron structure in electron-deuteron scattering.
With an isoscalar target, the assumption of charge symmetry, and in a region dominated by valence quarks, the contributions from 
PDFs in Eqn.~\ref{eq:A_beam} cancel in ratio and $A_\mathrm{PV}^\mathrm{electron}$ is proportional to
$\frac{20}{3} \sin^2 \theta_W - 1$. Thus, by PV asymmetry measurements in the valence region using longitudinally polarized electrons
and unpolarized deuterons, one can access the weak mixing angle.  

The available kinematic regime at the EIC, in particular the range $100 < Q^2 < 5000 $ GeV$^2$, has only scarcely been explored for measurements of $\sin^2\theta_W$~\cite{Kumar:2013yoa}. Therefore this measurement is sensitive to
testing the SM predictions and consequently, possible search for new physics beyond the SM, such as hypothesized dark-Z bosons~\cite{Davoudiasl:2015bua}. 
In such a new physics search, 
the effective model-independent weak coupling constants $C_{1q}$, $C_{2q}$ are used instead to describe the axial-vector and vector-axial 
couplings of electrons and quarks. At leading order of one-photon and one-$Z$ exchanges, they correspond to \cite{Wang:2014guo}
\begin{eqnarray}
C_{1u} &=& 2 g_A^e g_V^u, ~~ C_{2u} = 2 g_V^e g_A^u, \\
C_{1d} &=& 2 g_A^e g_V^d, ~~ C_{2d} = 2 g_V^e g_A^d,
\end{eqnarray}
where $g_A$ and $g_V$ are axial and vector couplings of super-scripted particle.  Higher-order radiative corrections must ultimately 
be taken into account, but are precisely predicted within the SM.  Any deviation from the predictions provides information on new 
contact interactions beyond the SM \cite{Wang:2014bba,Buckley:2012tc,GonzalezAlonso:2012jb}.

\section{Description of Simulations}
\label{simulation}
The event generator package DJANGOH \cite{Charchula:1994kf} which simulates deep inelastic lepton-proton (nuclei) 
scattering including both QED and QCD radiative effects, was employed for the simulation. 
DJANGOH is an interface of the Monte Carlo programs HERACLES \cite{Kwiatkowski:1990es} and LEPTO \cite{Ingelman:1996mq}. 
The HERACLES generator treats the $ep$ scattering using either parametrized structure functions 
or PDFs in the framework of the quark-parton model. The 
LEPTO program can integrate electroweak cross sections and simulate lepton-nucleon scattering 
with hadronic final states by using the JETSET library \cite{Sjostrand:1986hx}.
The DJANGOH package was used at HERA for studies with unpolarized proton beams and since then 
has been updated to include longitudinal polarization of the hadron beam which was used for the 
simulation study of weak charged current DIS at the a future EIC \cite{Aschenauer:2013iia}.
In our study, only the scattered electrons from the output of DJANGOH are investigated. The CTEQ61M~\cite{Stump:2003yu}
PDFs were used to evaluate the unpolarized structure functions 
while the DSSV08~\cite{deFlorian:2008mr} PDFs were used for the polarized structure function evaluations.

Electromagnetic radiation and finite detector resolution will induce bin migrations amongst events, which 
were studied and corrected in our analysis. Detector response to the generated events was simulated by 
a series of smearing parameters according to the studied prescription of conceptual designs of sPHENIX~\cite{sPHENIX,Adare:2015kwa} and ePHENIX~\cite{Adare:2014aaa} detector collaborations. In this procedure the 
lab-observables $\theta$, $\phi$, and momentum of outgoing electrons are convoluted with a Gaussian 
distribution and then the kinematic variables are recalculated. The standard deviation used in the smearing
are tabulated in Table \ref{table:smearing}. For the barrel (-1.1$<$$\eta$$<$1.1), the sPHENIX design~\cite{sPHENIX,Adare:2015kwa}
was used, while the ePHENIX design~\cite{Adare:2014aaa} was used in the $|\eta|>1.1$ region.
\begin{table}
\begin{center}
\begin{tabular}{|c|c|c|}\hline
       &     -1.1$<$ $\eta$ $<$1.1              &    $|\eta|>1.1$    \\  
       &      sPHENIX                           &     ePHENIX    \\ \hline
$d\theta$ (mrad)                 &      10                                 &      1           \\ \hline
$d\phi$ (mrad)                   &      0.3                                &      0.3          \\ \hline
$\frac{dp_{T}}{p_{T}}$          &      0.65\% $\oplus$ 0.09\% * $p_{T}$        &   0.65\%  $\oplus$  1\% * $p_{T}$           \\ \hline
$\frac{dE}{E}$                  &     3\% $\oplus$ 11.7\%/$\sqrt{E}$           &   1\%   $\oplus$ 2.5\%/$\sqrt{E}$         \\ \hline
\end{tabular}
\caption{Resolution parameters for detector smearing of scattered electrons. Quadrature sums are denoted by``$\oplus$". 
The sPHENIX~\cite{sPHENIX,Adare:2015kwa} was used as the barrel detector (-1.1$<$ $\eta$ $<$1.1), while the 
ePHENIX~\cite{Adare:2014aaa} was used for the $|\eta|>1.1$ region.}
\label{table:smearing}
\end{center}
\end{table}

Electron-proton collisions were used to study the $\gamma-Z$ interference structure functions and electron-deuteron
collisions were used to study the sensitivity to the weak-mixing angle. The data were analyzed in two-dimensional bins spaced logarithmically in $Q^2$ and $x$.
For the structure functions study, the data were further binned in $y$ in order to carry out $y$-dependent fits to 
extract projections on individual structure functions: $F_1^{\gamma Z}$ and $F_3^{\gamma Z}$ for the polarized electron beam, $g_1^{\gamma Z}$ 
and $g_5^{\gamma Z}$ for the polarized proton beam. In the following sensitivity study, PDF evolution and higher-twist effect 
were assumed to be small. Since the PVDIS asymmetry is proportional to $Q^2$, a weighted analysis was performed. In
each bin, a log-likelihood is defined as
\begin{equation}
\mathcal{L}=-\log \prod_\mathrm{events} \left(1-\lambda P f(Q^2) A\right) \sigma_{0},
\end{equation}
where 
$\sigma_{0}$ is the unpolarized cross section, $P$ is the polarization of the electron or proton beam, 
$\lambda = \pm1$ is the helicity states of incident electron or proton spin directions relative to the electron beam direction, 
$A_{PV}^{electron}$ and $A_{PV}^{hadron}$ have been written as $f(Q^2)A$, 
$f(Q^2)= \frac{G_F Q^2}{2 \sqrt{2} \pi \alpha}$,
$A=g_A^e \frac{F_1^{\gamma Z}}{F_1^{\gamma}} +  g_V^e \frac{Y_{-}}{2Y_{+}} \frac{F_3^{\gamma Z}}{F_1^{\gamma}}$
for the polarized electron beam or $A=g_V^e \frac{g_5^{\gamma Z}}{F_1^{\gamma}} + g_A^e \frac{Y_{-}}{Y_{+}} \frac{g_1^{\gamma Z}}{F_1^{\gamma}}$ for
the polarized proton beam. To minimize the log-likelihood, one finds the uncertainty of $A$ to be 
\begin{equation}
\sigma_A = \sqrt{ \frac{1}{\sum_\mathrm{events} \lambda^2 P^2 f^2(Q^2)} }.
\end{equation}
In each ($Q^2$, x) bin, the $\sigma_A$ was calculated in several $y$ bins, and a $y$-dependent fit was performed to 
extract the uncertainties for structure functions.

Toward extraction of uncertainties on the weak mixing angle, the $A_{PV}^{electron}$ on a deuteron target is written as
\begin{equation}
A_{PV}^{electron}=f(Q^2)[ f_1(Q^2,x) + f_2(Q^2, x)\sin^2 \theta_W],
\end{equation}
where 
\begin{equation}
\begin{split}
f_1(Q^2,x)= &- \frac{\frac{1}{6}(u+\bar{u})+\frac{1}{12}(d+\bar{d})+\frac{1}{12}(s+\bar{s})}{F_1^{\gamma}} \\
 &- \frac{Y_{-}}{4Y_{+}} \frac{F_3^{\gamma Z}}{F_1^{\gamma}},
\end{split}
\end{equation}
\begin{equation}
\begin{split}
f_2(Q^2,x)=&\frac{\frac{4}{9}(u+\bar{u})+\frac{1}{9}(d+\bar{d}) + \frac{1}{9}(s+\bar{s})}{F_1^{\gamma}} \\ 
&+ \frac{Y_{-}}{Y_{+}} \frac{F_3^{\gamma Z}}{F_1^{\gamma}}.
\end{split}
\end{equation}
Within a specific bin, a log-likelihood is defined as
\begin{equation}
\begin{split}
\mathcal{L}=-\log \prod_\mathrm{events} [ &1 - \lambda P f(Q^2) \times  \\
&(f_1(Q^2,x) + f_2(Q^2, x)\sin^2 \theta_W ) ] \sigma_{0}.
\end{split}
\end{equation}
The uncertainty in $\sin^2 \theta_W$ that minimizes the log-likelihood is
\begin{equation}
\sigma_{\sin^2 \theta_W } = \sqrt{ \frac{1}{\sum_{events} \lambda^2 P^2 f^2(Q^2) f_2^2(Q^2,x) } }.
\end{equation}

The basic kinematic cuts $1~\mathrm{GeV}^2 < Q^2 < (80~\mathrm{GeV})^2$, $W_h > 2~\mathrm{GeV}$ were used 
to select events in deep inelastic region.
Additionally the cut $y>0.1$ and electron momentum cut $P > 2$~GeV ($3$~GeV) were used for the 10~GeV (15~GeV) electron beam to ensure robust electron identification. 
The momentum cut on electrons only affects the statistics in the low $x$ region of the proposed measurements.
For the $\sin^2 \theta_W$ study, a cut of $x>0.2$ was imposed for electron-deuteron collisions to ensure 
that the contribution from PDF uncertainties is negligible.

Kinematic unfolding was performed for kinematic migration due to the effects of radiation and finite detector resolution. The instantaneous luminosity was assumed to be $10^{34} /\mathrm{cm}^2/\mathrm{s}$ per nucleon.
After considering detector efficiency of 70\% and beam-delivery efficiency of 70\%, this luminosity results in
an integrated sample 40~$\mathrm{fb}^{-1}/$month. A data collection period
for electron-proton collision to be over 2.5 years with 5 months of running per year is assumed, 
yielding a total integrated luminosity of 500~fb$^{-1}$.  The electron beam polarization is assumed to 
be 80\% while the proton beam polarization to be 70\% \cite{Accardi:2012qut}. For electron-deuteron collision, a
200 days of dedicated run is proposed, which corresponds to 267~$\mathrm{fb}^{-1}$ of integrated luminosity. It is noted that no deuteron polarization is required for the measurement of $\sin^2 \theta_W$.

\begin{figure}
\begin{center}
\resizebox{0.50\textwidth}{!}{ \includegraphics{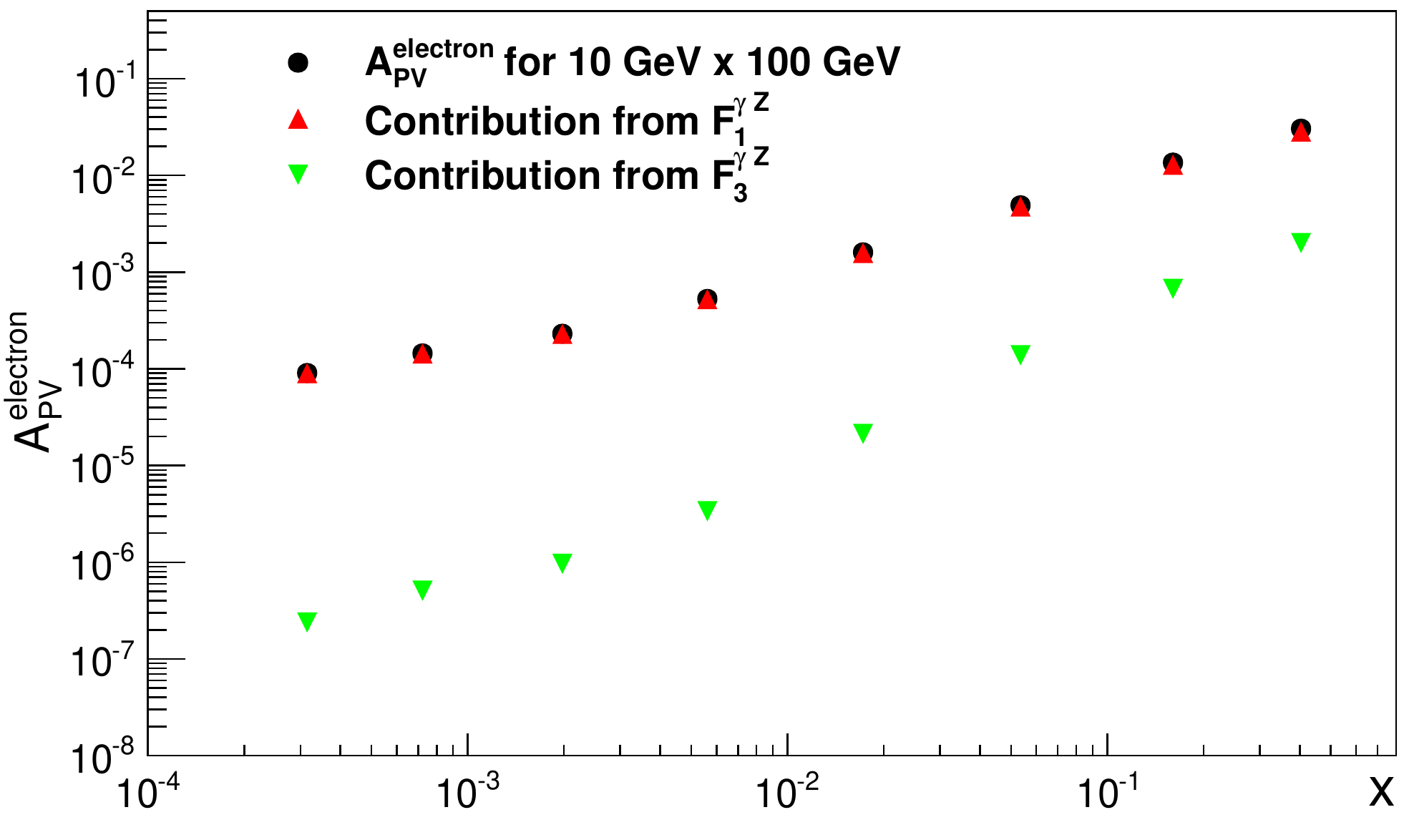} }
\caption{(Color Online) The predicted asymmetry $A_\mathrm{PV}^\mathrm{electron}$ vs. $x$ with 10~GeV longitudinally polarized electron on 100~GeV unpolarized proton collisions at an EIC.}
\label{fig:unpol_A}
\end{center}
\end{figure}

\begin{figure}
\begin{center}
\resizebox{0.50\textwidth}{!}{ \includegraphics{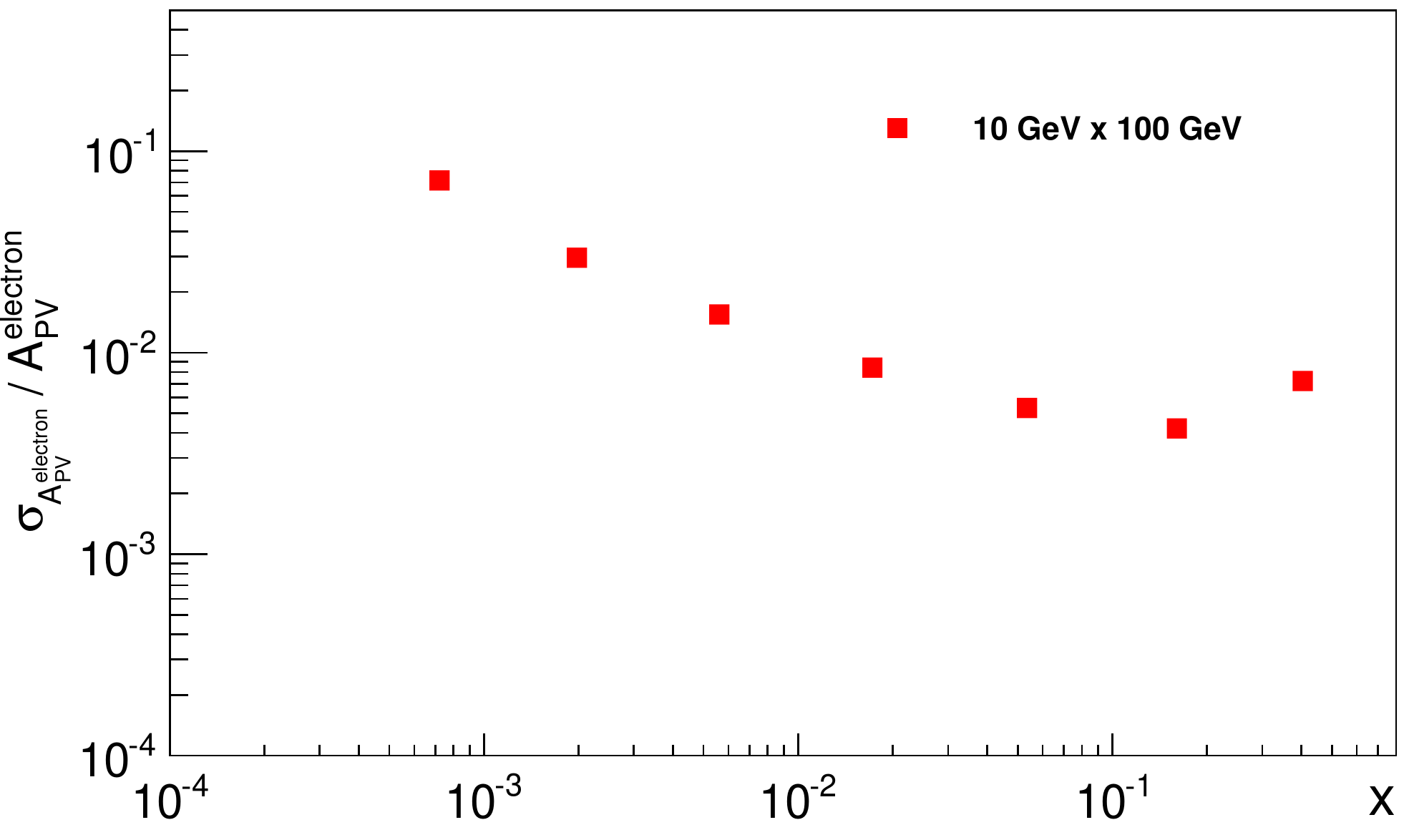} }
\caption{(Color Online) The predicted relative uncertainty for the measured asymmetry vs. $x$ with 10~GeV longitudinally polarized electron on 100~GeV unpolarized proton. The integrated luminosity of 500~$\mathrm{fb}^{-1}$ and electron beam polarization of 80\% are assumed.}
\label{fig:unpol_dA_over_A}
\end{center}
\end{figure}

\begin{figure}
\begin{center}
\resizebox{0.50\textwidth}{!}{ \includegraphics{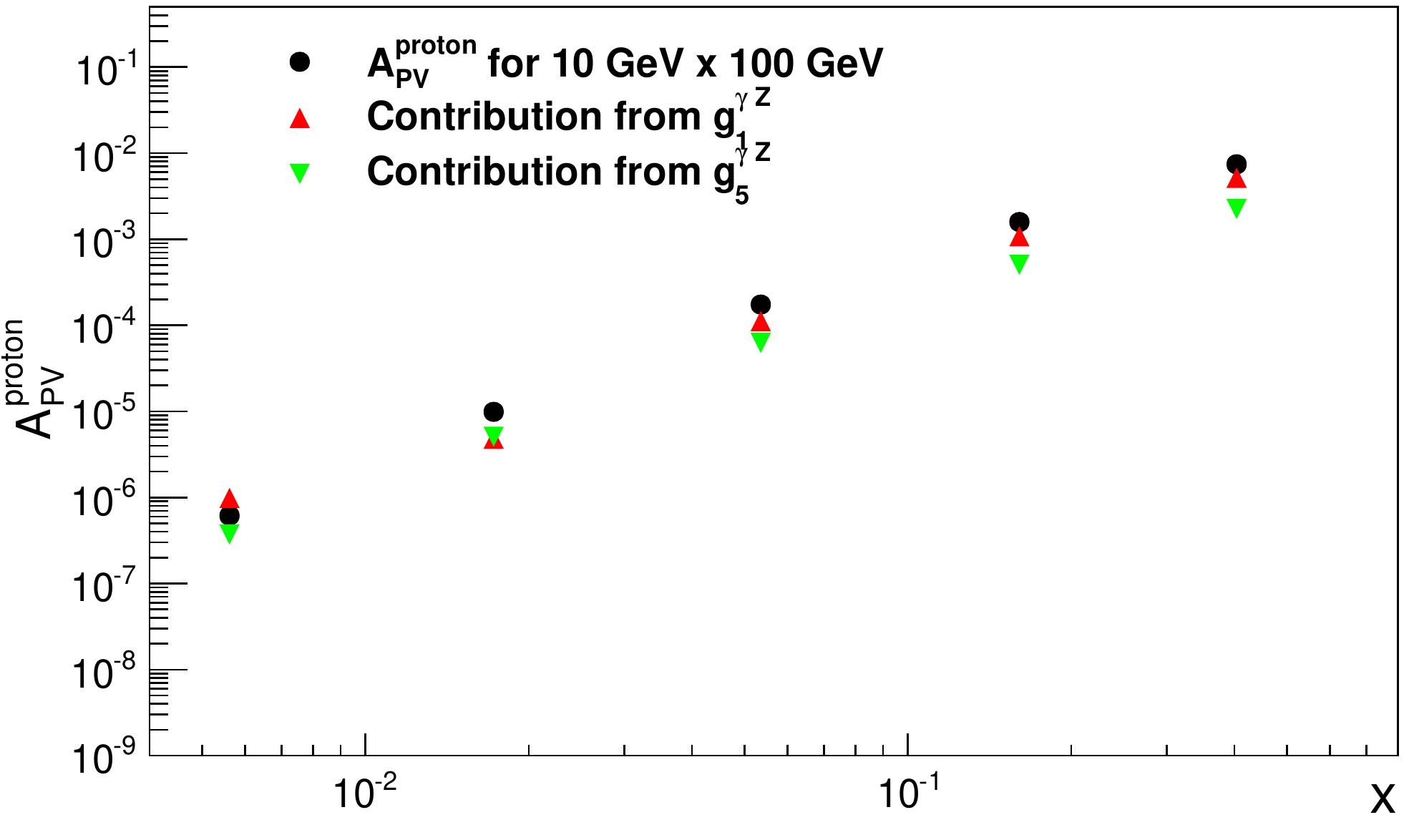} }
\caption{(Color Online) The predicted asymmetry $A_\mathrm{PV}^\mathrm{proton}$ vs. $x$ with 10~GeV unpolarized electron on 100~GeV longitudinally polarized proton.}
\label{fig:pol_A}
\end{center}
\end{figure}

\begin{figure}
\begin{center}
\resizebox{0.50\textwidth}{!}{ \includegraphics{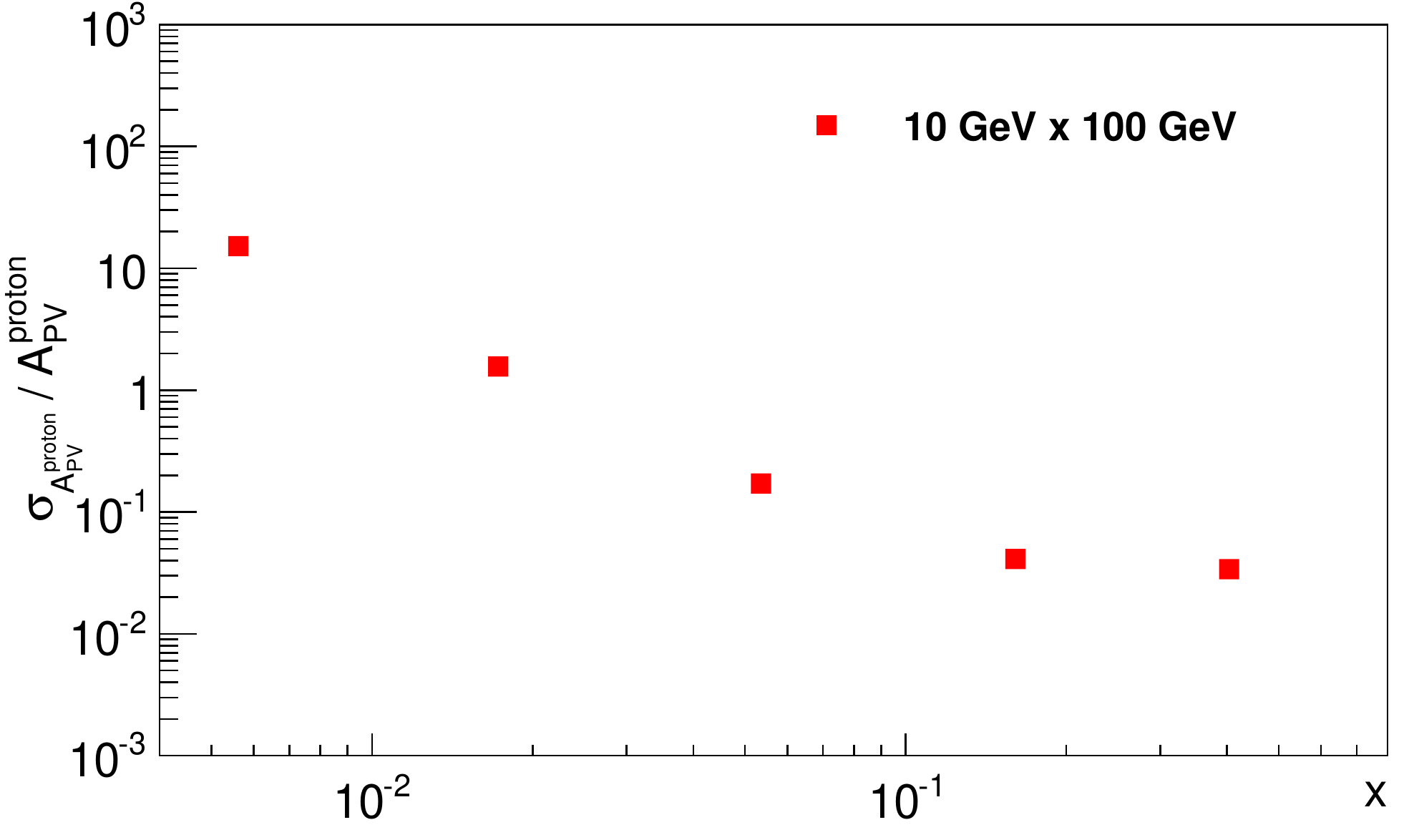} }
\caption{(Color Online) The predicted relative uncertainty for the measured asymmetry vs. $x$ with 10~GeV unpolarized electron on 100~GeV longitudinally polarized proton. The integrated luminosity of 500~$\mathrm{fb}^{-1}$ and 
proton beam polarization of 70\% are assumed.}
\label{fig:pol_dA_over_A}
\end{center}
\end{figure}

For the structure functions study, four different beam energy configurations were used: 10~GeV (electron) 
$\times$ 100 GeV (proton),  10~GeV $\times$ 250~GeV, 15~GeV $\times$ 100~GeV, 15~GeV $\times$ 250~GeV. As an example, the predicted $A_\mathrm{PV}^\mathrm{electron}$ as well as individual 
contributions from $F_1^{\gamma Z}$ and $F_3^{\gamma Z}$ terms are shown in Figure \ref{fig:unpol_A} 
for 10~GeV electron beam on 100~GeV proton beam. The data in different $Q^2$ bins for the same $x$ bin are 
combined statistically so that the plot is shown as a function of $x$. The measured relative uncertainty of the $A_\mathrm{PV}^\mathrm{electron}$ with assumed luminosity and beam polarization 
is shown in Figure~\ref{fig:unpol_dA_over_A}. In order to control the systematic uncertainty to be sub-dominant 
when compared with the statistical uncertainty, the differential luminosity systematic uncertainty should be 
controlled to be less than $10^{-4}$ at a future EIC. In polarized proton-proton collision at RHIC, it has already 
been achieved at the level of $10^{-4}$ recently \cite{Adare:2015ozj}. The predictions for $A_\mathrm{PV}^\mathrm{proton}$ using polarized proton beam are shown in Figures \ref{fig:pol_A} and \ref{fig:pol_dA_over_A}.  

\begin{figure}
\begin{center}
\resizebox{0.50\textwidth}{!}{ \includegraphics{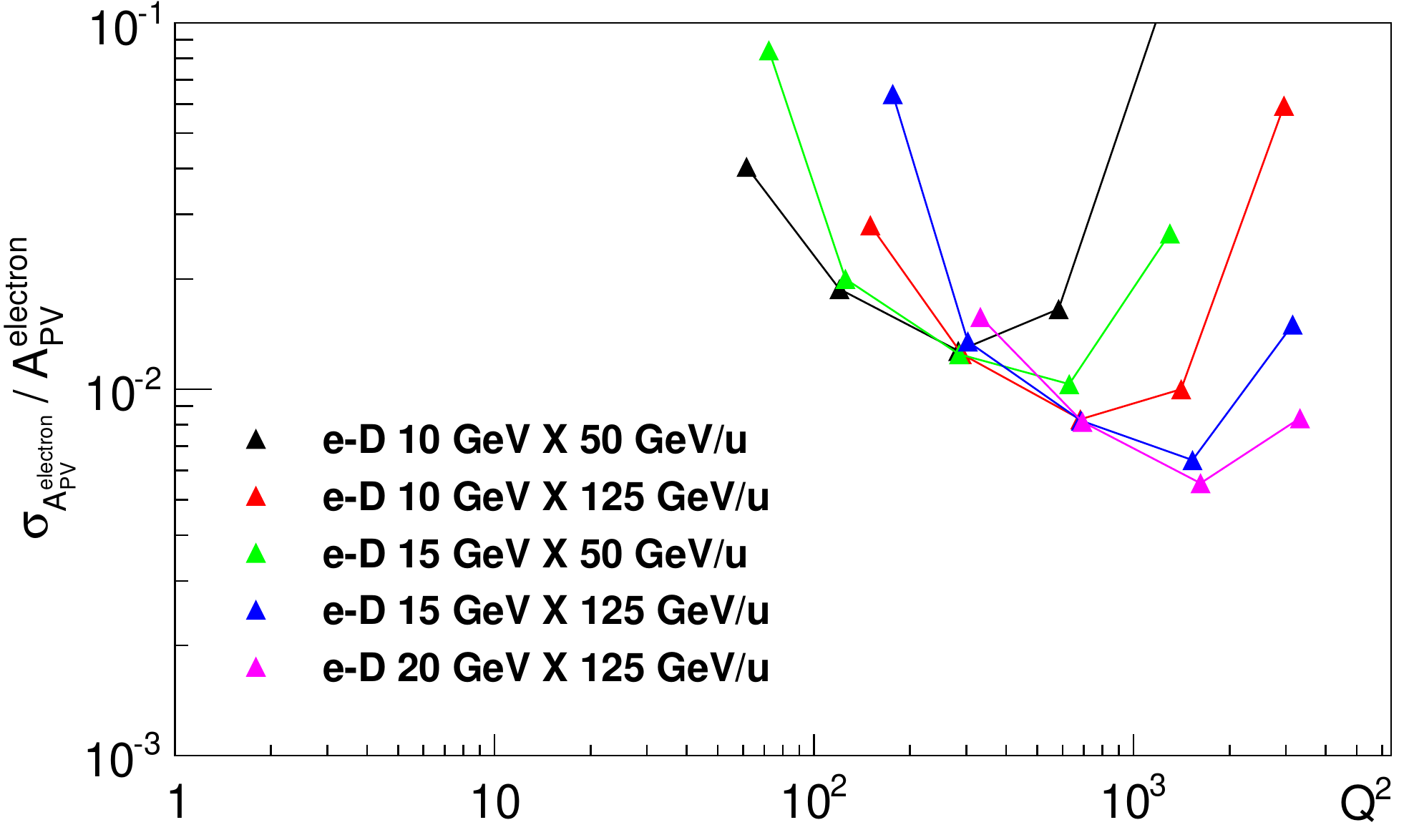} }
\caption{(Color Online) The predicted relative uncertainty on the measured asymmetry vs. $Q^{2}$ for electron-Deuteron collisions. The integrated luminosity of 267~$\mathrm{fb}^{-1}$ and the electron beam polarization 80\% are assumed.}
\label{fig:sin_dA_over_A}
\end{center}
\end{figure}

For the weak-mixing angle study, five different beam energy configurations were used for electron-deuteron collisions: 
10~GeV (electron) $\times$ 50 GeV (per nucleon), 10~GeV $\times$ 125~GeV, 15~GeV $\times$ 50~GeV, 15~GeV $\times$ 125~GeV, 20~GeV $\times$ 125~GeV.
The measured relative
uncertainties of $A_\mathrm{PV}^\mathrm{electron}$ as a function of $Q^2$ for different beam energy configurations are shown in 
Figure \ref{fig:sin_dA_over_A}.

\section{Results}
\label{sec:results}
\begin{figure}
\begin{center}
\resizebox{0.5\textwidth}{!}{\includegraphics{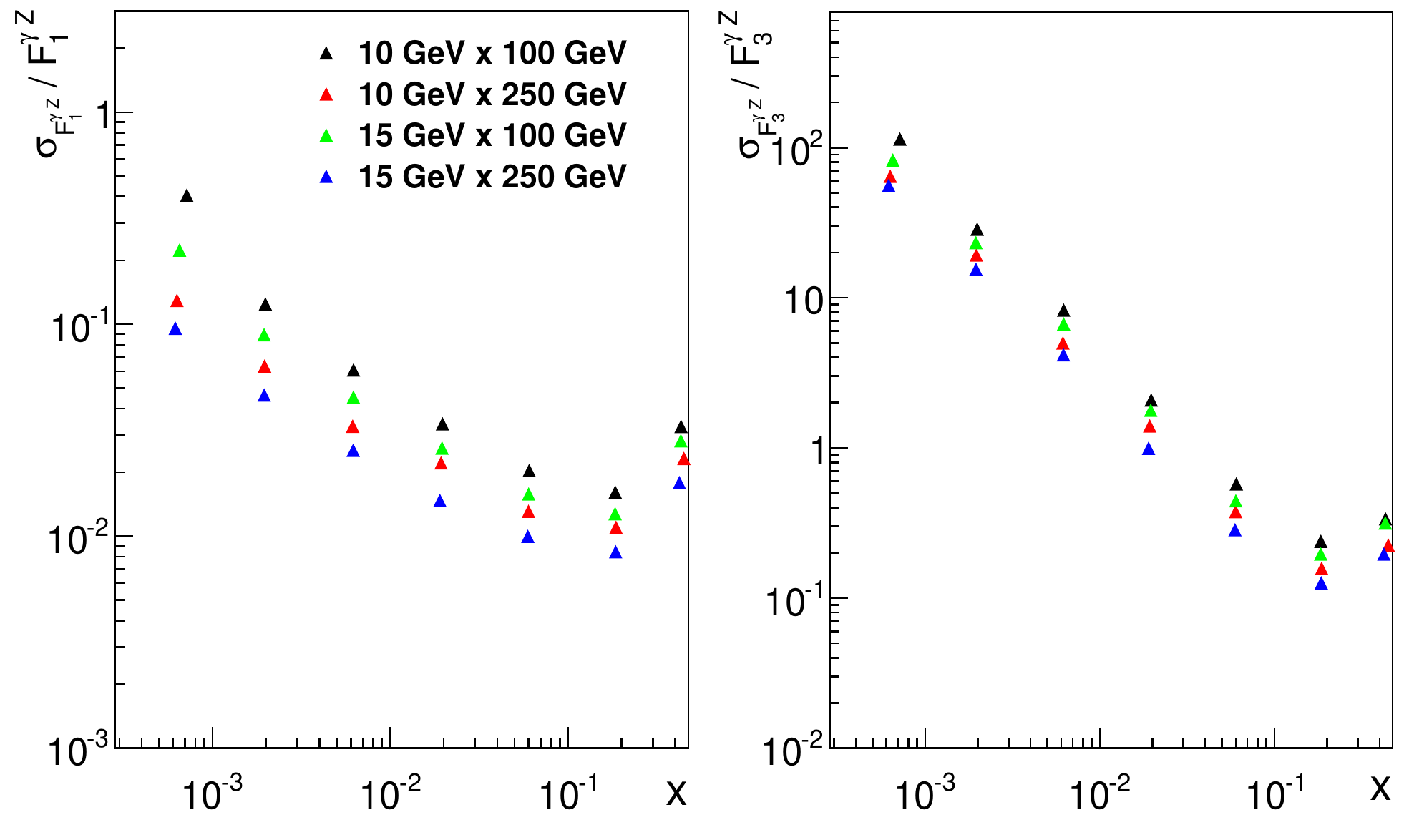} }
\caption{(Color Online) The projected relative uncertainties on $F_1^{\gamma Z}$ and $F_3^{\gamma Z}$ vs. $x$ from $A_\mathrm{PV}^\mathrm{electron}$ calculated after unfolding for bin migrations due to radiation and finite detector resolution using electron-proton collisions assuming an integrated luminosity of 500~$\mathrm{fb}^{-1}$. The electron polarization of
80\% was assumed.}
\label{fig:F_projection}
\end{center}
\end{figure}

\begin{figure}
\begin{center}
\resizebox{0.5\textwidth}{!}{\includegraphics{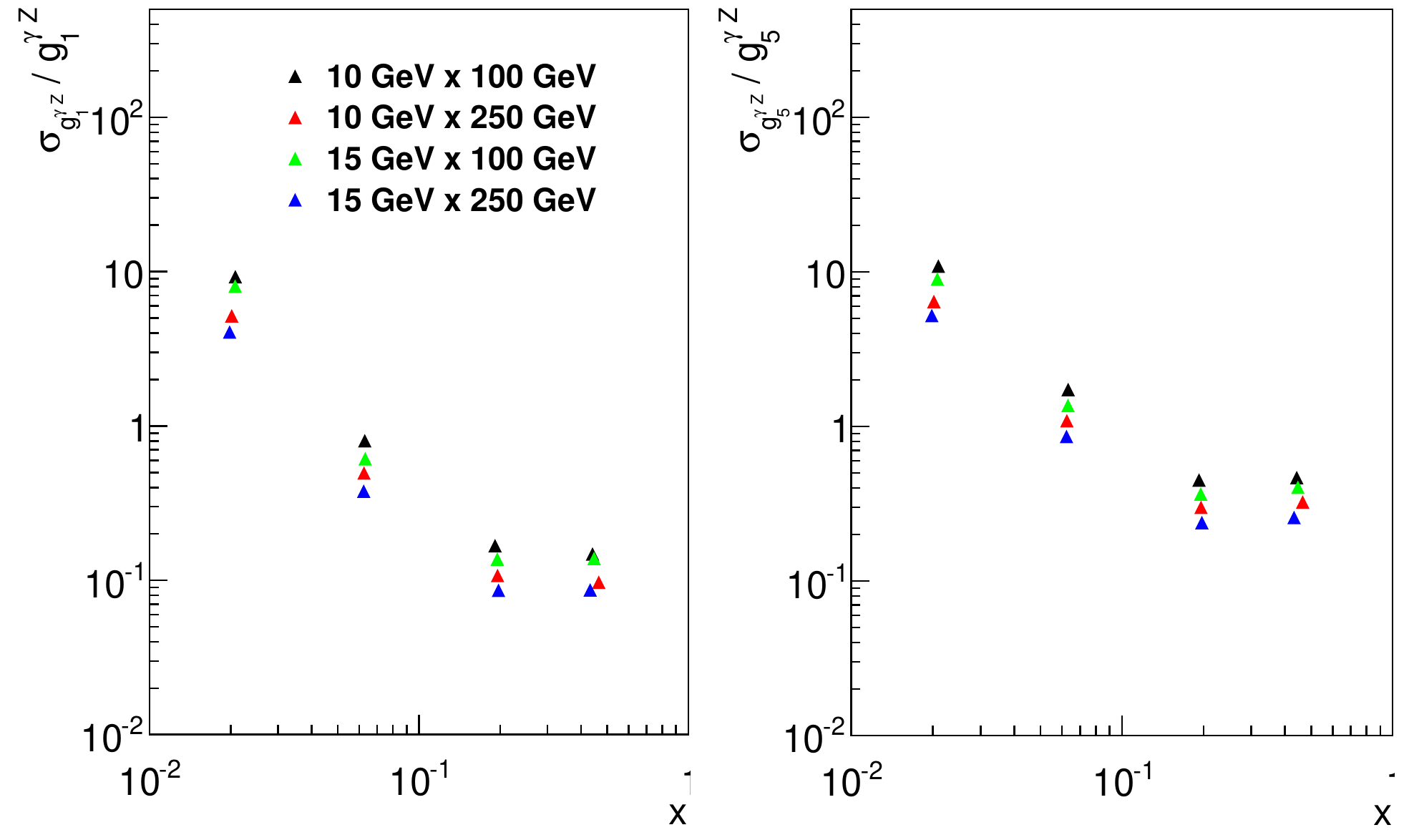} }
\caption{(Color Online) The projected relative uncertainties on $g_1^{\gamma Z}$ and $g_5^{\gamma Z}$ vs. $x$ from $A_\mathrm{PV}^{proton}$ 
calculated after unfolding of bin migrations due to radiation and finite detector resolution using electron-proton
collisions with an integrated luminosity of 500~$\mathrm{fb}^{-1}$ and the proton polarization of 70\%.}
\label{fig:g_projection}
\end{center}
\end{figure}

The projections for unpolarized $\gamma-Z$ interference structure functions
are shown as a function of $x$ in Figure \ref{fig:F_projection} with the data in different $Q^2$ bins combined statistically. The projections along with the $x, Q^2$ binning information and the predicted
value for the structure functions for different beam energy combinations are summarized in Tables~\ref{table_F_projection_10_100}, 
\ref{table_F_projection_10_250}, \ref{table_F_projection_15_100} and \ref{table_F_projection_15_250}. 
The projections for polarized $\gamma-Z$ interference structure functions are presented
in Figure \ref{fig:g_projection}. The values for different beam energy combinations are presented in Tables \ref{table_g_projection_10_100}, \ref{table_g_projection_10_250},  
\ref{table_g_projection_15_100} and \ref{table_g_projection_15_250}. Note that the mean values of the numbers shown 
in the tables are the $f^2(Q^2)$-weighted values after all the cuts are made (discussed in  Sec.~\ref{simulation}). The projections on asymmetries before the extraction of structure functions through $y$-dependent fits are also available. 

\begin{figure*}
\begin{center}
\resizebox{0.95\textwidth}{!}{\includegraphics{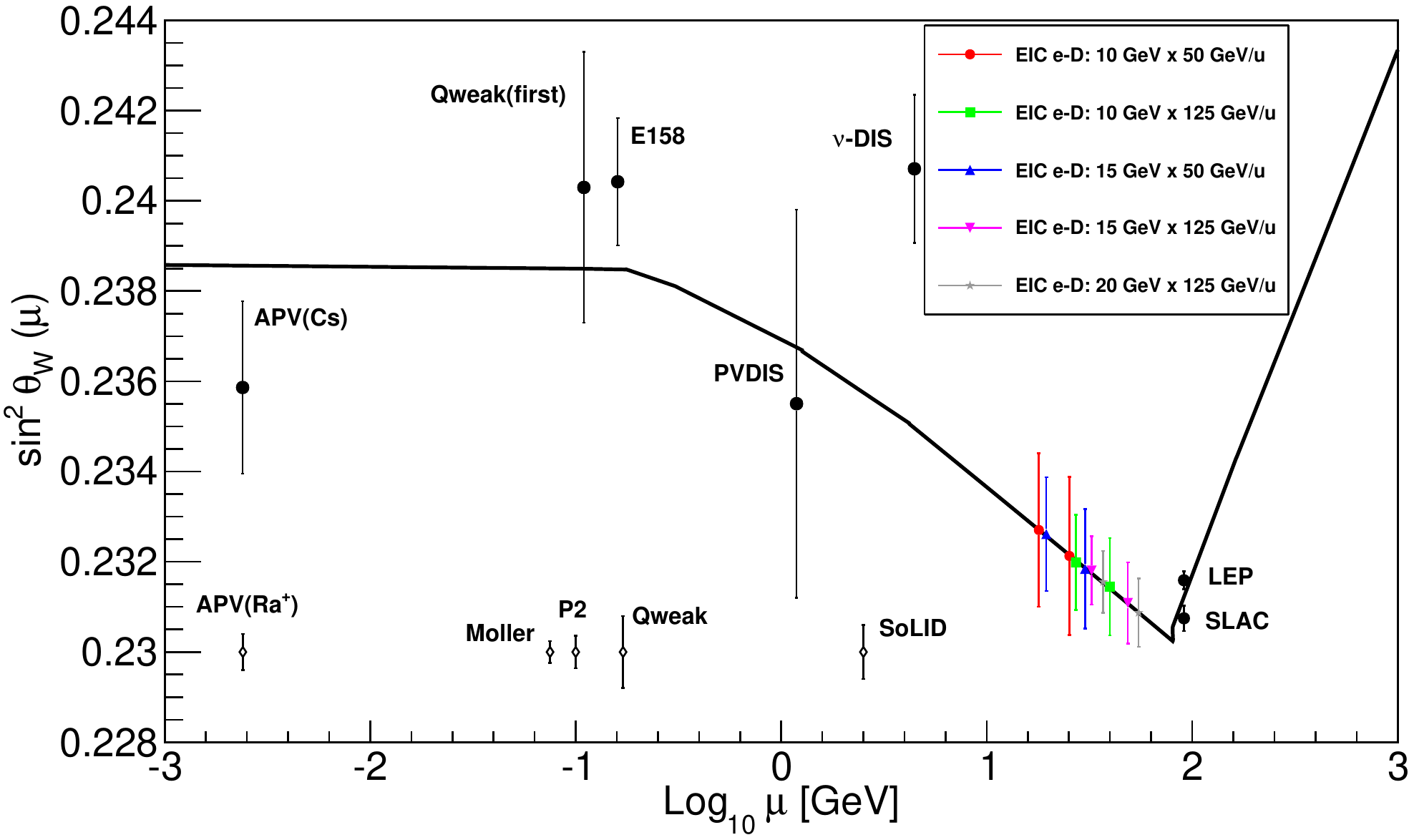} }
\caption{(Color Online) The projected uncertainties on the weak mixing angle vs. average $\mu {\rm or} Q$ of the measurement are shown at the appropriate average $\mu$ (or $Q$)
values for the integrated luminosity of 267~$\mathrm{fb}^{-1}$ per nucleon in electron-deuteron collisions for different 
energy configurations at an EIC. The existing measurements and other projected determinations at lower $\mu$ anticipated over the next 
decade are also shown for comparison. The scale dependence of the weak mixing angle is defined in the modified minimal subtraction scheme 
($\overline{\rm{MS}}$ scheme) \cite{Erler:2004in}.}
\label{fig:sin_world}
\end{center}
\end{figure*}

The projections of uncertainties on the weak 
mixing angle are shown along with existing or planed measurements at their appropriate average $\mu$ (or $Q$) 
values in Figure \ref{fig:sin_world}. The data points as well as the scale dependence of the weak mixing 
angle are defined in the modified minimal subtraction scheme ($\overline{\rm{MS}}$ scheme) \cite{Erler:2004in}. 
For each beam energy configuration at an EIC, the $Q^2$ coverage is divided into two regions. Within each region, the projections 
on the weak mixing angle were combined with appropriate statistical weights. Hence, two data points are shown for each beam energy configuration 
with the listed luminosity and electron beam polarization in the Figure \ref{fig:sin_world}.

\section{Summary and Outlook}
\label{summary}
A detailed phenomenological study of unpolarized and polarized electroweak
interference structure functions as well as weak mixing angle in neutral current mediated DIS
off protons and deuterons at a future EIC has been performed. The simulations were based on the event generator
package DJANGOH and conceptual detector design of sPHENIX and ePHENIX. The study has taken into account the corrections of
QED radiation of scattered electrons at next-to-leading order accuracy and bin migrations due to finite
detector resolutions. Different beam energy configurations, under discussion for a  future EIC
have been investigated.

The $\gamma-Z$ interference structure functions provide unique combinations of unpolarized and polarized PDFs in the parton model. Moreover, they have direct sensitivity to unpolarized and polarized strange quark
distributions. Along with the charged-current
mediated structure functions~\cite{Aschenauer:2013iia}, these structure functions could be very impactful input for
a clean extraction of individual PDFs. The combined measurements also provide an opportunity to test SU(3) flavor symmetry.
The study shows that higher center-of-mass with high luminosity is favorable for such extraction. The major systematic uncertainty of such
measurements stems from the uncertainties in the measurements of the polarization of the electron and proton beams. 
The requirement on the accuracy of electron (proton) beam polarimeters is $<1\%$ ($<3\%$).
A recent combined analysis with unpolarized data from both H1 and ZEUS at HERA has showed a slightly better 
precision on the $F_3^{\gamma Z}$ measurement \cite{Abramowicz:2015mha}.
However, the study proposed at an EIC would be far more powerful in 
constraining $F_1^{\gamma Z}$ since it is the dominant contribution to 
the parity-violating asymmetry.

The measurements of the weak mixing angle accessible at a future EIC are in a unique $Q^2$ region where there are no proposed
measurements in the following decade. Pioneering measurements in this region were carried out by HERA.
A combined QCD analysis on the weak mixing angle at HERA covers a broad high $Q^2$ region \cite{Abramowicz:2016ztw}, while the precision is significantly lower in the
$Q^2$ region covered by the proposed EIC measurements.
The impact of the measurements will depend on the status of searches for 
physics beyond the Standard Model. There
could be growing interest in such measurements depending on the outcomes of new physics searches at the LHC and elsewhere.

Armed with these results, a comprehensive study on PDF fits is planned for both unpolarized and polarized distributions. The study will
be focused on the impact on individual PDFs when combining data of different world data subsets with EIC projections. It might
be interesting to know how well the $s$ and $\Delta s$ distributions could be constrained without using semi-inclusive
measurements. Another interesting topic is the impact of
the improved unpolarized PDFs to LHC physics with EIC data.

In summary, a future EIC, with its high energy and high luminosity, opens up a new window for the study of neutral current electroweak physics.
New unpolarized and polarized $\gamma-Z$ interference structure functions can be accessed with relatively high precision, which provides important
inputs to achieve deeper understanding of nucleon structure. The measurements of the weak mixing angle in a unique kinematic region, far beyond 
the reach of the fixed target program and stretching towards the $Z^0$ pole probed at LEP and SLAC, could be of renewed interest 
in the following decade.

\begin{table*}
\begin{center}
\begin{tabular}{|c|c|c|c|c|c|c|c|c|}\hline
 $\log_{10}(Q^2)$ bin  &       $\log_{10}(x)$ bin &       $\langle Q^2 \rangle$ (GeV$^2$) &      $\langle x \rangle$  & $y$ coverage    &  $\frac{\sigma_{F_3^{\gamma Z}}}{F_3^{\gamma Z}}$ & $\langle F_3^{\gamma Z} \rangle$ & $\frac{\sigma_{F_1^{\gamma Z}}}{F_1^{\gamma Z}}$  & $\langle F_1^{\gamma Z} \rangle$ \\ \hline \hline
(0.0,	0.4)&	(-3.5,	-3.0)&	1.6e+00&     7.2e-04&     (0.25, 0.81)&	1.1e+02&     3.8e+01&	4.1e-01&     2.8e+02 \\	
(0.0,	0.4)&	(-3.0,	-2.5)&	1.8e+00&     1.9e-03&     (0.10, 0.63)&	5.8e+01&     2.6e+01&	1.8e-01&     9.8e+01 \\	
(0.0,	0.4)&	(-2.5,	-2.0)&	2.1e+00&     4.1e-03&     (0.10, 0.20)&	5.1e+02&     1.8e+01&	1.1e+00&     4.0e+01 \\	
(0.4,	0.8)&	(-3.5,	-3.0)&	2.8e+00&     9.2e-04&     (0.63, 0.81)&	2.0e+03&     3.6e+01&	9.4e+00&     2.5e+02 \\	
(0.4,	0.8)&	(-3.0,	-2.5)&	4.2e+00&     2.1e-03&     (0.20, 0.82)&	3.4e+01&     2.7e+01&	1.8e-01&     1.3e+02 \\	
(0.4,	0.8)&	(-2.5,	-2.0)&	4.6e+00&     5.8e-03&     (0.10, 0.50)&	2.8e+01&     1.8e+01&	1.1e-01&     4.4e+01 \\	
(0.4,	0.8)&	(-2.0,	-1.5)&	5.5e+00&     1.2e-02&     (0.10, 0.16)&	1.4e+03&     1.3e+01&	4.0e+00&     1.9e+01 \\	
(0.8,	1.2)&	(-3.0,	-2.5)&	7.6e+00&     2.7e-03&     (0.50, 0.83)&	1.5e+02&     2.6e+01&	1.1e+00&     1.2e+02 \\	
(0.8,	1.2)&	(-2.5,	-2.0)&	1.1e+01&     6.3e-03&     (0.16, 0.84)&	9.8e+00&     1.9e+01&	7.7e-02&     5.3e+01 \\	
(0.8,	1.2)&	(-2.0,	-1.5)&	1.2e+01&     1.7e-02&     (0.10, 0.40)&	1.5e+01&     1.2e+01&	8.0e-02&     1.6e+01 \\	
(0.8,	1.2)&	(-1.5,	-1.0)&	1.5e+01&     3.4e-02&     (0.10, 0.13)&	9.2e+03&     8.4e+00&	4.7e+01&     6.6e+00 \\	
(1.2,	1.6)&	(-2.5,	-2.0)&	2.2e+01&     8.1e-03&     (0.40, 0.89)&	1.9e+01&     1.8e+01&	2.5e-01&     4.2e+01 \\	
(1.2,	1.6)&	(-2.0,	-1.5)&	2.8e+01&     1.9e-02&     (0.13, 0.90)&	3.3e+00&     1.2e+01&	4.3e-02&     1.7e+01 \\	
(1.2,	1.6)&	(-1.5,	-1.0)&	3.0e+01&     5.0e-02&     (0.10, 0.31)&	8.1e+00&     7.3e+00&	7.2e-02&     4.9e+00 \\	
(1.6,	2.0)&	(-2.0,	-1.5)&	6.4e+01&     2.3e-02&     (0.32, 1.0)&	2.8e+00&     1.1e+01&	7.3e-02&     1.4e+01 \\	
(1.6,	2.0)&	(-1.5,	-1.0)&	6.9e+01&     5.9e-02&     (0.10, 0.79)&	1.1e+00&     6.8e+00&	2.7e-02&     4.4e+00 \\	
(1.6,	2.0)&	(-1.0,	-0.5)&	7.8e+01&     1.4e-01&     (0.10, 0.25)&	6.7e+00&     3.5e+00&	9.9e-02&     1.3e+00 \\	
(2.0,	2.4)&	(-2.0,	-1.5)&	1.1e+02&     2.9e-02&     (0.79, 1.0)&	8.3e+01&     9.9e+00&	3.2e+00&     9.7e+00 \\	
(2.0,	2.4)&	(-1.5,	-1.0)&	1.6e+02&     6.4e-02&     (0.25, 1.0)&	6.8e-01&     6.5e+00&	3.5e-02&     3.9e+00 \\	
(2.0,	2.4)&	(-1.0,	-0.5)&	1.7e+02&     1.8e-01&     (0.10, 0.63)&	6.7e-01&     2.8e+00&	2.4e-02&     9.6e-01 \\	
(2.0,	2.4)&	(-0.5,	0.0)&	2.0e+02&     3.8e-01&     (0.10, 0.20)&	2.3e+01&     6.8e-01&	4.4e-01&     1.7e-01 \\	
(2.4,	2.8)&	(-1.5,	-1.0)&	3.0e+02&     8.5e-02&     (0.63, 1.0)&	4.7e+00&     5.1e+00&	3.7e-01&     2.4e+00 \\	
(2.4,	2.8)&	(-1.0,	-0.5)&	4.2e+02&     1.8e-01&     (0.20, 1.0)&	2.6e-01&     2.6e+00&	2.2e-02&     8.5e-01 \\	
(2.4,	2.8)&	(-0.5,	0.0)&	4.4e+02&     4.2e-01&     (0.10, 0.49)&	1.1e+00&     5.3e-01&	5.3e-02&     1.3e-01 \\	
(2.8,	3.2)&	(-1.0,	-0.5)&	8.2e+02&     2.5e-01&     (0.50, 1.0)&	1.0e+00&     1.5e+00&	1.3e-01&     4.3e-01 \\	
(2.8,	3.2)&	(-0.5,	0.0)&	1.1e+03&     4.2e-01&     (0.20, 1.0)&	3.5e-01&     4.6e-01&	4.1e-02&     1.1e-01 \\	
(3.2,	3.6)&	(-0.5,	0.0)&	1.9e+03&     5.3e-01&     (0.54, 1.0)&	2.1e+00&     1.7e-01&	3.2e-01&     3.9e-02 \\ \hline	
\end{tabular}
\caption{
Anticipated sensitivities of $F_1^{\gamma Z}$ and $F_3^{\gamma Z}$ functions for individual bins in the ($Q^2$, $x$) plane. 
The projections are for the 10 GeV longitudinally polarized electron beam on the 100 GeV unpolarized proton beam. 
The first two columns define the ($\log_{10}(Q^2)$, $\log_{10}(x)$) two dimensional bins. 
The $\langle Q^2 \rangle$ and $\langle x \rangle$ are the $f^2(Q^2)$ weighted (as discussed in Sec. \ref{simulation}) mean values in each bin.
The $y$ coverage for the bin is also tabulated.
The $\langle F_1^{\gamma Z} \rangle$ and $\langle F_3^{\gamma Z} \rangle$ are the predicted mean values for the structure functions.
The $\frac{\sigma_{F_1^{\gamma Z}}}{F_1^{\gamma Z}}$ and  $\frac{\sigma_{F_3^{\gamma Z}}}{F_3^{\gamma Z}}$ are the projected relative uncertainties.
The cuts mentioned in Sec. \ref{simulation} are applied to the data.
}
\label{table_F_projection_10_100}
\end{center}
\end{table*}

\begin{table*}
\begin{center}
\begin{tabular}{|c|c|c|c|c|c|c|c|c|}\hline
      $\log_{10}(Q^2)$ bin  &           $\log_{10}(x)$ bin &    $\langle Q^2 \rangle$ GeV$^2$&     $\langle x \rangle$   & $y$ coverage  &  $\frac{\sigma_{F_3^{\gamma Z}}}{F_3^{\gamma Z}}$ & $\langle F_3^{\gamma Z} \rangle$    &  $\frac{\sigma_{F_1^{\gamma Z}}}{F_1^{\gamma Z}}$   & $\langle F_1^{\gamma Z} \rangle$ \\  \hline  \hline
(0.0,	0.4)&	(-4.0,	-3.5)&	1.5e+00&     2.4e-04&     (0.32, 0.81)&	3.4e+02&     5.9e+01&	5.6e-01&     1.0e+03 \\	
(0.0,	0.4)&	(-3.5,	-3.0)&	1.8e+00&     6.0e-04&     (0.10, 0.79)&	9.6e+01&     4.2e+01&	1.6e-01&     3.9e+02 \\	
(0.0,	0.4)&	(-3.0,	-2.5)&	2.0e+00&     1.4e-03&     (0.10, 0.25)&	3.7e+02&     2.9e+01&	4.5e-01&     1.3e+02 \\	
(0.4,	0.8)&	(-4.0,	-3.5)&	2.5e+00&     3.1e-04&     (0.79, 0.81)&	7.4e+06&     5.7e+01&	1.6e+04&     8.9e+02 \\	
(0.4,	0.8)&	(-3.5,	-3.0)&	4.0e+00&     7.2e-04&     (0.25, 0.82)&	8.7e+01&     4.4e+01&	2.3e-01&     4.4e+02 \\	
(0.4,	0.8)&	(-3.0,	-2.5)&	4.5e+00&     1.9e-03&     (0.10, 0.63)&	4.2e+01&     3.0e+01&	9.1e-02&     1.6e+02 \\	
(0.4,	0.8)&	(-2.5,	-2.0)&	5.2e+00&     4.1e-03&     (0.10, 0.20)&	4.1e+02&     2.1e+01&	6.3e-01&     6.3e+01 \\	
(0.8,	1.2)&	(-3.5,	-3.0)&	6.9e+00&     9.2e-04&     (0.63, 0.82)&	1.3e+03&     4.2e+01&	4.5e+00&     4.1e+02 \\	
(0.8,	1.2)&	(-3.0,	-2.5)&	1.0e+01&     2.1e-03&     (0.20, 0.84)&	2.3e+01&     3.1e+01&	9.1e-02&     1.9e+02 \\	
(0.8,	1.2)&	(-2.5,	-2.0)&	1.1e+01&     5.7e-03&     (0.10, 0.50)&	1.9e+01&     2.0e+01&	5.9e-02&     6.1e+01 \\	
(0.8,	1.2)&	(-2.0,	-1.5)&	1.4e+01&     1.2e-02&     (0.10, 0.16)&	9.6e+02&     1.4e+01&	2.4e+00&     2.4e+01 \\	
(1.2,	1.6)&	(-3.0,	-2.5)&	2.0e+01&     2.7e-03&     (0.50, 0.87)&	8.5e+01&     2.9e+01&	5.2e-01&     1.6e+02 \\	
(1.2,	1.6)&	(-2.5,	-2.0)&	2.7e+01&     6.1e-03&     (0.16, 0.90)&	5.9e+00&     2.1e+01&	4.2e-02&     6.8e+01 \\	
(1.2,	1.6)&	(-2.0,	-1.5)&	2.9e+01&     1.7e-02&     (0.10, 0.40)&	9.4e+00&     1.3e+01&	4.7e-02&     1.9e+01 \\	
(1.2,	1.6)&	(-1.5,	-1.0)&	3.7e+01&     3.4e-02&     (0.10, 0.13)&	2.3e+03&     8.7e+00&	1.2e+01&     7.2e+00 \\	
(1.6,	2.0)&	(-2.5,	-2.0)&	6.1e+01&     7.9e-03&     (0.40, 1.0)&	1.0e+01&     1.9e+01&	1.3e-01&     5.2e+01 \\	
(1.6,	2.0)&	(-2.0,	-1.5)&	7.0e+01&     1.9e-02&     (0.13, 1.0)&	2.4e+00&     1.3e+01&	3.1e-02&     2.0e+01 \\	
(1.6,	2.0)&	(-1.5,	-1.0)&	7.5e+01&     4.9e-02&     (0.10, 0.32)&	5.1e+00&     7.5e+00&	4.4e-02&     5.2e+00 \\	
(2.0,	2.4)&	(-2.0,	-1.5)&	1.6e+02&     2.2e-02&     (0.32, 1.0)&	1.8e+00&     1.2e+01&	4.3e-02&     1.6e+01 \\	
(2.0,	2.4)&	(-1.5,	-1.0)&	1.7e+02&     5.8e-02&     (0.10, 0.79)&	7.4e-01&     6.9e+00&	1.7e-02&     4.6e+00 \\	
(2.0,	2.4)&	(-1.0,	-0.5)&	1.9e+02&     1.4e-01&     (0.10, 0.25)&	5.4e+00&     3.4e+00&	7.7e-02&     1.2e+00 \\	
(2.4,	2.8)&	(-2.0,	-1.5)&	2.8e+02&     2.9e-02&     (0.80, 1.0)&	5.5e+01&     1.0e+01&	2.0e+00&     1.0e+01 \\	
(2.4,	2.8)&	(-1.5,	-1.0)&	4.1e+02&     6.3e-02&     (0.25, 1.0)&	4.4e-01&     6.5e+00&	2.3e-02&     4.1e+00 \\	
(2.4,	2.8)&	(-1.0,	-0.5)&	4.4e+02&     1.7e-01&     (0.10, 0.63)&	4.6e-01&     2.8e+00&	1.6e-02&     9.4e-01 \\	
(2.4,	2.8)&	(-0.5,	0.0)&	5.1e+02&     3.8e-01&     (0.10, 0.20)&	2.4e+01&     6.7e-01&	4.5e-01&     1.7e-01 \\	
(2.8,	3.2)&	(-1.5,	-1.0)&	7.6e+02&     8.5e-02&     (0.64, 1.0)&	3.1e+00&     5.1e+00&	2.5e-01&     2.4e+00 \\	
(2.8,	3.2)&	(-1.0,	-0.5)&	1.1e+03&     1.8e-01&     (0.20, 1.0)&	1.7e-01&     2.5e+00&	1.5e-02&     8.4e-01 \\	
(2.8,	3.2)&	(-0.5,	0.0)&	1.1e+03&     4.2e-01&     (0.10, 0.49)&	8.3e-01&     4.9e-01&	4.0e-02&     1.2e-01 \\	
(3.2,	3.6)&	(-1.0,	-0.5)&	2.1e+03&     2.5e-01&     (0.51, 1.0)&	5.6e-01&     1.4e+00&	7.0e-02&     4.0e-01 \\	
(3.2,	3.6)&	(-0.5,	0.0)&	2.7e+03&     4.2e-01&     (0.20, 1.0)&	2.2e-01&     4.4e-01&	2.7e-02&     1.1e-01 \\	
(3.6,	4.0)&	(-0.5,	0.0)&	5.1e+03&     5.7e-01&     (0.59, 1.0)&	1.2e+00&     1.1e-01&	1.9e-01&     2.6e-02 \\ \hline	
\end{tabular}
\caption{
Anticipated sensitivities of $F_1^{\gamma Z}$ and $F_3^{\gamma Z}$ functions for individual bins in the ($Q^2$, $x$) plane.
The projections are for the 10 GeV longitudinally polarized electron beam on the 250 GeV unpolarized proton beam.
The first two columns define the ($\log_{10}(Q^2)$, $\log_{10}(x)$) two dimensional bins.
The $\langle Q^2 \rangle$ and $\langle x \rangle$ are the $f^2(Q^2)$ weighted (as discussed in Sec. \ref{simulation}) mean values in each bin.
The $y$ coverage for the bin is also tabulated.
The $\langle F_1^{\gamma Z} \rangle$ and $\langle F_3^{\gamma Z} \rangle$ are the predicted mean values for the structure functions.
The $\frac{\sigma_{F_1^{\gamma Z}}}{F_1^{\gamma Z}}$ and  $\frac{\sigma_{F_3^{\gamma Z}}}{F_3^{\gamma Z}}$ are the projected relative uncertainties.
The cuts mentioned in Sec. \ref{simulation} are applied to the data.
}
\label{table_F_projection_10_250}
\end{center}
\end{table*}

\begin{table*}
\begin{center}
\begin{tabular}{|c|c|c|c|c|c|c|c|c|}\hline
       $\log_{10}(Q^2)$ bin  &            $\log_{10}(x)$ bin  &  $\langle Q^2 \rangle$ (GeV$^2$)&    $\langle x \rangle$   &  $y$ coverage  &  $\frac{\sigma_{F_3^{\gamma Z}}}{F_3^{\gamma Z}}$ & $\langle F_3^{\gamma Z} \rangle$    &  $\frac{\sigma_{F_1^{\gamma Z}}}{F_1^{\gamma Z}}$   & $\langle F_1^{\gamma Z} \rangle$ \\  \hline  \hline
(0.0,	0.4)&	(-4.0,	-3.5)&	1.2e+00&     2.8e-04&     (0.53, 0.80)&	1.7e+03&     5.5e+01&	3.5e+00&     8.1e+02 \\	
(0.0,	0.4)&	(-3.5,	-3.0)&	1.7e+00&     6.4e-04&     (0.17, 0.80)&	8.8e+01&     4.0e+01&	2.3e-01&     3.4e+02 \\	
(0.0,	0.4)&	(-3.0,	-2.5)&	1.9e+00&     1.8e-03&     (0.10, 0.42)&	1.0e+02&     2.6e+01&	2.0e-01&     1.1e+02 \\	
(0.0,	0.4)&	(-2.5,	-2.0)&	2.3e+00&     3.5e-03&     (0.10, 0.13)&	4.2e+03&     2.0e+01&	7.1e+00&     4.9e+01 \\	
(0.4,	0.8)&	(-3.5,	-3.0)&	3.3e+00&     8.3e-04&     (0.42, 0.81)&	2.5e+02&     3.9e+01&	9.3e-01&     3.2e+02 \\	
(0.4,	0.8)&	(-3.0,	-2.5)&	4.4e+00&     2.0e-03&     (0.13, 0.81)&	2.9e+01&     2.9e+01&	1.1e-01&     1.5e+02 \\	
(0.4,	0.8)&	(-2.5,	-2.0)&	4.8e+00&     5.2e-03&     (0.10, 0.33)&	5.9e+01&     1.9e+01&	1.4e-01&     5.0e+01 \\	
(0.4,	0.8)&	(-2.0,	-1.5)&	6.2e+00&     1.0e-02&     (0.10, 0.11)&	4.4e+04&     1.4e+01&	1.1e+02&     2.3e+01 \\	
(0.8,	1.2)&	(-3.0,	-2.5)&	9.2e+00&     2.5e-03&     (0.33, 0.82)&	4.3e+01&     2.8e+01&	2.5e-01&     1.4e+02 \\	
(0.8,	1.2)&	(-2.5,	-2.0)&	1.1e+01&     6.0e-03&     (0.11, 0.82)&	1.0e+01&     2.0e+01&	5.6e-02&     5.8e+01 \\	
(0.8,	1.2)&	(-2.0,	-1.5)&	1.2e+01&     1.5e-02&     (0.10, 0.26)&	3.9e+01&     1.3e+01&	1.4e-01&     1.9e+01 \\	
(1.2,	1.6)&	(-2.5,	-2.0)&	2.5e+01&     7.2e-03&     (0.26, 0.84)&	9.0e+00&     1.9e+01&	9.2e-02&     5.2e+01 \\	
(1.2,	1.6)&	(-2.0,	-1.5)&	2.8e+01&     1.8e-02&     (0.10, 0.66)&	4.1e+00&     1.2e+01&	3.6e-02&     1.7e+01 \\	
(1.2,	1.6)&	(-1.5,	-1.0)&	3.2e+01&     4.1e-02&     (0.10, 0.21)&	2.7e+01&     7.9e+00&	1.8e-01&     5.8e+00 \\	
(1.6,	2.0)&	(-2.5,	-2.0)&	4.4e+01&     9.3e-03&     (0.67, 0.86)&	1.5e+02&     1.7e+01&	2.3e+00&     3.9e+01 \\	
(1.6,	2.0)&	(-2.0,	-1.5)&	6.5e+01&     2.1e-02&     (0.21, 0.91)&	2.1e+00&     1.2e+01&	4.0e-02&     1.7e+01 \\	
(1.6,	2.0)&	(-1.5,	-1.0)&	7.1e+01&     5.7e-02&     (0.10, 0.53)&	1.9e+00&     6.9e+00&	2.7e-02&     4.6e+00 \\	
(1.6,	2.0)&	(-1.0,	-0.5)&	8.6e+01&     1.2e-01&     (0.10, 0.17)&	2.9e+01&     3.9e+00&	3.5e-01&     1.5e+00 \\	
(2.0,	2.4)&	(-2.0,	-1.5)&	1.3e+02&     2.7e-02&     (0.53, 1.0)&	6.4e+00&     1.0e+01&	2.1e-01&     1.1e+01 \\	
(2.0,	2.4)&	(-1.5,	-1.0)&	1.7e+02&     5.9e-02&     (0.17, 1.0)&	5.1e-01&     6.9e+00&	2.0e-02&     4.5e+00 \\	
(2.0,	2.4)&	(-1.0,	-0.5)&	1.8e+02&     1.6e-01&     (0.10, 0.42)&	1.2e+00&     3.0e+00&	2.7e-02&     1.0e+00 \\	
(2.0,	2.4)&	(-0.5,	0.0)&	2.3e+02&     3.5e-01&     (0.10, 0.13)&	9.7e+01&     8.2e-01&	1.6e+00&     2.1e-01 \\	
(2.4,	2.8)&	(-1.5,	-1.0)&	3.6e+02&     7.6e-02&     (0.42, 1.0)&	1.0e+00&     5.7e+00&	6.7e-02&     2.9e+00 \\	
(2.4,	2.8)&	(-1.0,	-0.5)&	4.4e+02&     1.8e-01&     (0.13, 1.0)&	2.4e-01&     2.7e+00&	1.5e-02&     9.3e-01 \\	
(2.4,	2.8)&	(-0.5,	0.0)&	4.5e+02&     4.1e-01&     (0.10, 0.33)&	2.1e+00&     5.6e-01&	6.5e-02&     1.4e-01 \\	
(2.8,	3.2)&	(-1.0,	-0.5)&	9.6e+02&     2.2e-01&     (0.34, 1.0)&	3.3e-01&     2.0e+00&	3.6e-02&     6.1e-01 \\	
(2.8,	3.2)&	(-0.5,	0.0)&	1.1e+03&     4.2e-01&     (0.13, 0.83)&	4.0e-01&     4.9e-01&	3.4e-02&     1.2e-01 \\	
(3.2,	3.6)&	(-1.0,	-0.5)&	1.7e+03&     3.0e-01&     (0.86, 1.0)&	2.9e+01&     1.0e+00&	4.2e+00&     2.6e-01 \\	
(3.2,	3.6)&	(-0.5,	0.0)&	2.2e+03&     4.7e-01&     (0.34, 1.0)&	4.9e-01&     3.3e-01&	7.0e-02&     7.9e-02 \\ \hline	
\end{tabular}
\caption{
Anticipated sensitivities of $F_1^{\gamma Z}$ and $F_3^{\gamma Z}$ functions for individual bins in the ($Q^2$, $x$) plane.
The projections are for the 15 GeV longitudinally polarized electron beam on the 100 GeV unpolarized proton beam.
The first two columns define the ($\log_{10}(Q^2)$, $\log_{10}(x)$) two dimensional bins.
The $\langle Q^2 \rangle$ and $\langle x \rangle$ are the $f^2(Q^2)$ weighted (as discussed in Sec. \ref{simulation}) mean values in each bin.
The $y$ coverage for the bin is also tabulated.
The $\langle F_1^{\gamma Z} \rangle$ and $\langle F_3^{\gamma Z} \rangle$ are the predicted mean values for the structure functions.
The $\frac{\sigma_{F_1^{\gamma Z}}}{F_1^{\gamma Z}}$ and  $\frac{\sigma_{F_3^{\gamma Z}}}{F_3^{\gamma Z}}$ are the projected relative uncertainties.
The cuts mentioned in Sec. \ref{simulation} are applied to the data.
}
\label{table_F_projection_15_100}
\end{center}
\end{table*}

\begin{table*}
\begin{center}
\begin{tabular}{|c|c|c|c|c|c|c|c|c|}\hline
      $\log_{10}(Q^2)$ bin  &      $\log_{10}(x)$ bin &   $\langle Q^2 \rangle$ (GeV$^2$) &    $\langle x \rangle$    &  $y$ coverage  &  $\frac{\sigma_{F_3^{\gamma Z}}}{F_3^{\gamma Z}}$ & $\langle F_3^{\gamma Z} \rangle$   &  $\frac{\sigma_{F_1^{\gamma Z}}}{F_1^{\gamma Z}}$   & $\langle F_1^{\gamma Z} \rangle$    \\      \hline  \hline
(0.0,	0.4)&	(-4.5,	-4.0)&	1.1e+00&     9.4e-05&     (0.67, 0.80)&	2.3e+04&     8.6e+01&	2.1e+01&     3.2e+03 \\	
(0.0,	0.4)&	(-4.0,	-3.5)&	1.6e+00&     2.1e-04&     (0.21, 0.80)&	2.2e+02&     6.4e+01&	2.7e-01&     1.3e+03 \\	
(0.0,	0.4)&	(-3.5,	-3.0)&	1.8e+00&     5.8e-04&     (0.10, 0.53)&	1.5e+02&     4.2e+01&	1.6e-01&     4.0e+02 \\	
(0.0,	0.4)&	(-3.0,	-2.5)&	2.2e+00&     1.2e-03&     (0.10, 0.17)&	1.9e+03&     3.1e+01&	1.7e+00&     1.6e+02 \\	
(0.4,	0.8)&	(-4.0,	-3.5)&	2.9e+00&     2.8e-04&     (0.53, 0.80)&	1.4e+03&     6.2e+01&	2.4e+00&     1.1e+03 \\	
(0.4,	0.8)&	(-3.5,	-3.0)&	4.3e+00&     6.5e-04&     (0.17, 0.81)&	6.4e+01&     4.7e+01&	1.2e-01&     5.5e+02 \\	
(0.4,	0.8)&	(-3.0,	-2.5)&	4.7e+00&     1.8e-03&     (0.10, 0.42)&	7.3e+01&     3.1e+01&	1.0e-01&     1.8e+02 \\	
(0.4,	0.8)&	(-2.5,	-2.0)&	5.8e+00&     3.5e-03&     (0.10, 0.13)&	2.8e+03&     2.3e+01&	3.5e+00&     7.7e+01 \\	
(0.8,	1.2)&	(-3.5,	-3.0)&	8.2e+00&     8.3e-04&     (0.42, 0.81)&	1.7e+02&     4.5e+01&	4.6e-01&     5.0e+02 \\	
(0.8,	1.2)&	(-3.0,	-2.5)&	1.1e+01&     1.9e-03&     (0.13, 0.82)&	1.9e+01&     3.3e+01&	5.7e-02&     2.2e+02 \\	
(0.8,	1.2)&	(-2.5,	-2.0)&	1.2e+01&     5.1e-03&     (0.10, 0.33)&	4.0e+01&     2.1e+01&	7.9e-02&     6.8e+01 \\	
(0.8,	1.2)&	(-2.0,	-1.5)&	1.6e+01&     1.0e-02&     (0.10, 0.11)&	2.4e+04&     1.5e+01&	5.1e+01&     2.8e+01 \\	
(1.2,	1.6)&	(-3.0,	-2.5)&	2.3e+01&     2.5e-03&     (0.33, 0.84)&	2.7e+01&     3.1e+01&	1.3e-01&     1.9e+02 \\	
(1.2,	1.6)&	(-2.5,	-2.0)&	2.8e+01&     5.9e-03&     (0.11, 0.84)&	6.8e+00&     2.1e+01&	3.2e-02&     7.3e+01 \\	
(1.2,	1.6)&	(-2.0,	-1.5)&	3.1e+01&     1.4e-02&     (0.10, 0.27)&	2.5e+01&     1.4e+01&	8.5e-02&     2.2e+01 \\	
(1.6,	2.0)&	(-3.0,	-2.5)&	4.0e+01&     3.2e-03&     (0.84, 0.84)&	7.6e+05&     2.9e+01&	5.1e+03&     1.5e+02 \\	
(1.6,	2.0)&	(-2.5,	-2.0)&	6.3e+01&     7.0e-03&     (0.27, 0.91)&	5.4e+00&     2.0e+01&	4.9e-02&     6.5e+01 \\	
(1.6,	2.0)&	(-2.0,	-1.5)&	7.0e+01&     1.8e-02&     (0.10, 0.67)&	2.7e+00&     1.3e+01&	2.1e-02&     2.0e+01 \\	
(1.6,	2.0)&	(-1.5,	-1.0)&	8.1e+01&     4.1e-02&     (0.10, 0.21)&	1.7e+01&     8.1e+00&	1.1e-01&     6.2e+00 \\	
(2.0,	2.4)&	(-2.5,	-2.0)&	1.1e+02&     9.0e-03&     (0.67, 0.96)&	5.0e+01&     1.8e+01&	7.1e-01&     4.7e+01 \\	
(2.0,	2.4)&	(-2.0,	-1.5)&	1.7e+02&     1.9e-02&     (0.21, 1.0)&	1.1e+00&     1.3e+01&	2.1e-02&     2.0e+01 \\	
(2.0,	2.4)&	(-1.5,	-1.0)&	1.8e+02&     5.6e-02&     (0.10, 0.53)&	1.2e+00&     7.1e+00&	1.7e-02&     4.7e+00 \\	
(2.0,	2.4)&	(-1.0,	-0.5)&	2.1e+02&     1.2e-01&     (0.10, 0.17)&	1.9e+01&     3.8e+00&	2.2e-01&     1.5e+00 \\	
(2.4,	2.8)&	(-2.0,	-1.5)&	3.3e+02&     2.6e-02&     (0.53, 1.0)&	3.7e+00&     1.1e+01&	1.2e-01&     1.3e+01 \\	
(2.4,	2.8)&	(-1.5,	-1.0)&	4.3e+02&     5.8e-02&     (0.17, 1.0)&	3.3e-01&     7.0e+00&	1.3e-02&     4.7e+00 \\	
(2.4,	2.8)&	(-1.0,	-0.5)&	4.5e+02&     1.6e-01&     (0.10, 0.42)&	8.4e-01&     2.9e+00&	1.9e-02&     1.0e+00 \\	
(2.4,	2.8)&	(-0.5,	0.0)&	5.7e+02&     3.5e-01&     (0.10, 0.13)&	7.0e+01&     8.2e-01&	1.2e+00&     2.1e-01 \\	
(2.8,	3.2)&	(-1.5,	-1.0)&	9.1e+02&     7.5e-02&     (0.42, 1.0)&	6.7e-01&     5.7e+00&	4.5e-02&     3.0e+00 \\	
(2.8,	3.2)&	(-1.0,	-0.5)&	1.1e+03&     1.7e-01&     (0.14, 1.0)&	1.6e-01&     2.7e+00&	1.0e-02&     9.2e-01 \\	
(2.8,	3.2)&	(-0.5,	0.0)&	1.1e+03&     4.1e-01&     (0.10, 0.33)&	1.4e+00&     5.5e-01&	4.4e-02&     1.4e-01 \\	
(3.2,	3.6)&	(-1.0,	-0.5)&	2.4e+03&     2.2e-01&     (0.34, 1.0)&	2.1e-01&     1.9e+00&	2.3e-02&     5.8e-01 \\	
(3.2,	3.6)&	(-0.5,	0.0)&	2.8e+03&     4.2e-01&     (0.15, 0.83)&	2.4e-01&     4.6e-01&	2.1e-02&     1.1e-01 \\	
(3.6,	4.0)&	(-1.0,	-0.5)&	4.3e+03&     2.9e-01&     (0.90, 1.0)&	4.8e+01&     9.6e-01&	7.0e+00&     2.5e-01 \\	
(3.6,	4.0)&	(-0.5,	0.0)&	5.5e+03&     4.5e-01&     (0.39, 1.0)&	3.4e-01&     3.4e-01&	4.8e-02&     8.3e-02 \\ \hline	
\end{tabular}
\caption{
Anticipated sensitivities of $F_1^{\gamma Z}$ and $F_3^{\gamma Z}$ functions for individual bins in the ($Q^2$, $x$) plane.
The projections are for the 15 GeV longitudinally polarized electron beam on the 250 GeV unpolarized proton beam.
The first two columns define the ($\log_{10}(Q^2)$, $\log_{10}(x)$) two dimensional bins.
The $\langle Q^2 \rangle$ and $\langle x \rangle$ are the $f^2(Q^2)$ weighted (as discussed in Sec. \ref{simulation}) mean values in each bin.
The $y$ coverage for the bin is also tabulated.
The $\langle F_1^{\gamma Z} \rangle$ and $\langle F_3^{\gamma Z} \rangle$ are the predicted mean values for the structure functions.
The $\frac{\sigma_{F_1^{\gamma Z}}}{F_1^{\gamma Z}}$ and  $\frac{\sigma_{F_3^{\gamma Z}}}{F_3^{\gamma Z}}$ are the projected relative uncertainties.
The cuts mentioned in Sec. \ref{simulation} are applied to the data.
}
\label{table_F_projection_15_250}
\end{center}
\end{table*}

\begin{table*}
\begin{center}
\begin{tabular}{|c|c|c|c|c|c|c|c|c|}\hline
      $\log_{10}(Q^2)$ bin  &    $\log_{10}(x)$ bin &  $\langle Q^2 \rangle$ (GeV$^2$)&     $\langle x \rangle$   & $y$ coverage  &  $\frac{\sigma_{g_1^{\gamma Z}}}{g_1^{\gamma Z}}$ & $\langle g_1^{\gamma Z} \rangle$    &  $\frac{\sigma_{g_5^{\gamma Z}}}{g_5^{\gamma Z}}$   & $\langle g_5^{\gamma Z} \rangle$   \\  \hline  \hline
(0.0,	0.4)	&(-3.5,	-3.0)	&1.6e+00&	7.2e-04&     (0.25, 0.81)&  -6.2e+01&     -3.0e+00&	5.4e+03&     3.2e-01 \\	
(0.0,	0.4)	&(-3.0,	-2.5)	&1.8e+00&	1.9e-03&     (0.10, 0.63)&  -5.1e+01&     -1.2e+00&	1.8e+03&     1.4e-01 \\	
(0.0,	0.4)	&(-2.5,	-2.0)	&2.1e+00&	4.1e-03&     (0.10, 0.20)&  -6.9e+02&     -4.7e-01&	3.6e+03&     1.5e-01 \\	
(0.4,	0.8)	&(-3.5,	-3.0)	&2.8e+00&	9.2e-04&     (0.63, 0.81)&  -1.4e+03&     -2.1e+00&	9.4e+04&     3.8e-01 \\	
(0.4,	0.8)	&(-3.0,	-2.5)	&4.2e+00&	2.1e-03&     (0.20, 0.82)&  -4.1e+01&     -9.7e-01&	9.7e+02&     3.5e-01 \\	
(0.4,	0.8)	&(-2.5,	-2.0)	&4.6e+00&	5.8e-03&     (0.10, 0.50)&  -7.3e+01&     -2.9e-01&	2.0e+02&     3.5e-01 \\	
(0.4,	0.8)	&(-2.0,	-1.5)	&5.5e+00&	1.2e-02&     (0.10, 0.16)&  -6.7e+03&     -3.9e-02&	9.3e+02&     4.4e-01 \\	
(0.8,	1.2)	&(-3.0,	-2.5)	&7.6e+00&	2.7e-03&     (0.50, 0.83)&  -2.5e+02&     -6.5e-01&	3.9e+03&     4.7e-01 \\	
(0.8,	1.2)	&(-2.5,	-2.0)	&1.1e+01&	6.3e-03&     (0.16, 0.84)&  -3.6e+01&     -2.2e-01&	1.2e+02&     5.2e-01 \\	
(0.8,	1.2)	&(-2.0,	-1.5)	&1.2e+01&	1.7e-02&     (0.10, 0.40)&  1.8e+02&     4.0e-02&	3.1e+01&     5.8e-01 \\	
(0.8,	1.2)	&(-1.5,	-1.0)	&1.5e+01&	3.4e-02&     (0.10, 0.13)&  2.1e+03&     1.3e-01&	6.3e+02&     6.6e-01 \\	
(1.2,	1.6)	&(-2.5,	-2.0)	&2.2e+01&	8.1e-03&     (0.40, 0.89)&  -1.6e+02&     -8.3e-02&	2.3e+02&     6.5e-01 \\	
(1.2,	1.6)	&(-2.0,	-1.5)	&2.8e+01&	1.9e-02&     (0.13, 0.90)&  1.9e+01&     7.8e-02&	1.5e+01&     7.0e-01 \\	
(1.2,	1.6)	&(-1.5,	-1.0)	&3.0e+01&	5.0e-02&     (0.10, 0.31)&  1.5e+01&     1.7e-01&	7.6e+00&     7.1e-01 \\	
(1.6,	2.0)	&(-2.0,	-1.5)	&6.4e+01&	2.3e-02&     (0.32, 1.0)&  1.0e+01&     1.3e-01&	1.9e+01&     7.9e-01 \\	
(1.6,	2.0)	&(-1.5,	-1.0)	&6.9e+01&	5.9e-02&     (0.10, 0.79)&  1.8e+00&     1.9e-01&	2.4e+00&     7.4e-01 \\	
(1.6,	2.0)	&(-1.0,	-0.5)	&7.8e+01&	1.4e-01&     (0.10, 0.25)&  5.2e+00&     1.8e-01&	3.1e+00&     5.6e-01 \\	
(2.0,	2.4)	&(-2.0,	-1.5)	&1.1e+02&	2.9e-02&     (0.79, 1.0)&  2.1e+02&     1.7e-01&	5.6e+02&     8.3e-01 \\	
(2.0,	2.4)	&(-1.5,	-1.0)	&1.6e+02&	6.4e-02&     (0.25, 1.0)&  9.1e-01&     2.0e-01&	2.7e+00&     7.6e-01 \\	
(2.0,	2.4)	&(-1.0,	-0.5)	&1.7e+02&	1.8e-01&     (0.10, 0.63)&  4.9e-01&     1.6e-01&	6.8e-01&     4.9e-01 \\	
(2.0,	2.4)	&(-0.5,	0.0)	&2.0e+02&	3.8e-01&     (0.10, 0.20)&  5.4e+00&     6.2e-02&	3.6e+00&     1.7e-01 \\	
(2.4,	2.8)	&(-1.5,	-1.0)	&3.0e+02&	8.5e-02&     (0.63, 1.0)&  4.9e+00&     2.1e-01&	1.9e+01&     7.1e-01 \\	
(2.4,	2.8)	&(-1.0,	-0.5)	&4.2e+02&	1.8e-01&     (0.20, 1.0)&  1.9e-01&     1.5e-01&	6.2e-01&     4.6e-01 \\	
(2.4,	2.8)	&(-0.5,	0.0)	&4.4e+02&	4.2e-01&     (0.10, 0.49)&  5.0e-01&     5.0e-02&	7.6e-01&     1.3e-01 \\	
(2.8,	3.2)	&(-1.0,	-0.5)	&8.2e+02&	2.5e-01&     (0.50, 1.0)&  5.9e-01&     1.1e-01&	2.6e+00&     3.2e-01 \\	
(2.8,	3.2)	&(-0.5,	0.0)	&1.1e+03&	4.2e-01&     (0.20, 1.0)&  1.5e-01&     4.4e-02&	5.9e-01&     1.2e-01 \\	
(3.2,	3.6)	&(-0.5,	0.0)	&1.9e+03&	5.3e-01&     (0.54, 1.0)&  7.8e-01&     1.9e-02&	3.8e+00&     5.0e-02 \\ \hline	
\end{tabular}
\caption{
Anticipated sensitivities of $g_1^{\gamma Z}$ and $g_5^{\gamma Z}$ functions for individual bins in the ($Q^2$, $x$) plane.
The projections are for the 10 GeV unpolarized electron beam on the 100 GeV longitudinally polarized proton beam.
The first two columns define the ($\log_{10}(Q^2)$, $\log_{10}(x)$) two dimensional bins.
The $\langle Q^2 \rangle$ and $\langle x \rangle$ are the $f^2(Q^2)$ weighted (as discussed in Sec. \ref{simulation}) mean values in each bin.
The $y$ coverage for the bin is also tabulated.
The $\langle g_1^{\gamma Z} \rangle$ and $\langle g_5^{\gamma Z} \rangle$ are the predicted mean values for the structure functions.
The $\frac{\sigma_{g_1^{\gamma Z}}}{g_1^{\gamma Z}}$ and  $\frac{\sigma_{g_5^{\gamma Z}}}{g_5^{\gamma Z}}$ are the projected relative uncertainties.
The cuts mentioned in Sec. \ref{simulation} are applied to the data.
}
\label{table_g_projection_10_100}
\end{center}
\end{table*}

\begin{table*}
\begin{center}
\begin{tabular}{|c|c|c|c|c|c|c|c|c|}\hline
     $\log_{10}(Q^2)$ bin  &      $\log_{10}(x)$ bin &    $\langle Q^2 \rangle$ (GeV$^2$) &    $\langle x \rangle$     & $y$ coverage &  $\frac{\sigma_{g_1^{\gamma Z}}}{g_1^{\gamma Z}}$ & $\langle g_1^{\gamma Z} \rangle$    &  $\frac{\sigma_{g_5^{\gamma Z}}}{g_5^{\gamma Z}}$   & $\langle g_5^{\gamma Z} \rangle$   \\      \hline  \hline
(0.0,	0.4)	&(-4.0,	-3.5)	&1.5e+00&	2.4e-04&     (0.32, 0.81)&  -1.1e+02&     -7.8e+00&	8.4e+03&     1.0e+00 \\	
(0.0,	0.4)	&(-3.5,	-3.0)	&1.8e+00&	6.0e-04&     (0.10, 0.79)&  -4.7e+01&     -3.6e+00&	2.0e+03&     4.5e-01 \\	
(0.0,	0.4)	&(-3.0,	-2.5)	&2.0e+00&	1.4e-03&     (0.10, 0.25)&  -3.0e+02&     -1.5e+00&	4.3e+03&     2.0e-01 \\	
(0.4,	0.8)	&(-4.0,	-3.5)	&2.5e+00&	3.1e-04&     (0.79, 0.81)&  -1.7e+06&     -5.8e+00&	1.3e+08&     9.2e-01 \\	
(0.4,	0.8)	&(-3.5,	-3.0)	&4.0e+00&	7.2e-04&     (0.25, 0.82)&  -5.8e+01&     -2.8e+00&	2.6e+03&     5.9e-01 \\	
(0.4,	0.8)	&(-3.0,	-2.5)	&4.5e+00&	1.9e-03&     (0.10, 0.63)&  -4.6e+01&     -1.1e+00&	5.5e+02&     3.9e-01 \\	
(0.4,	0.8)	&(-2.5,	-2.0)	&5.2e+00&	4.1e-03&     (0.10, 0.20)&  -6.6e+02&     -4.1e-01&	1.2e+03&     3.7e-01 \\	
(0.8,	1.2)	&(-3.5,	-3.0)	&6.9e+00&	9.2e-04&     (0.63, 0.82)&  -1.1e+03&     -2.1e+00&	4.3e+04&     6.4e-01 \\	
(0.8,	1.2)	&(-3.0,	-2.5)	&1.0e+01&	2.1e-03&     (0.20, 0.84)&  -3.2e+01&     -9.4e-01&	4.5e+02&     5.8e-01 \\	
(0.8,	1.2)	&(-2.5,	-2.0)	&1.1e+01&	5.7e-03&     (0.10, 0.50)&  -6.3e+01&     -2.5e-01&	9.6e+01&     5.4e-01 \\	
(0.8,	1.2)	&(-2.0,	-1.5)	&1.4e+01&	1.2e-02&     (0.10, 0.16)&  -5.5e+04&     -3.2e-03&	4.8e+02&     5.9e-01 \\	
(1.2,	1.6)	&(-3.0,	-2.5)	&2.0e+01&	2.7e-03&     (0.50, 0.87)&  -1.7e+02&     -6.3e-01&	1.8e+03&     6.8e-01 \\	
(1.2,	1.6)	&(-2.5,	-2.0)	&2.7e+01&	6.1e-03&     (0.16, 0.90)&  -2.7e+01&     -1.9e-01&	6.1e+01&     6.9e-01 \\	
(1.2,	1.6)	&(-2.0,	-1.5)	&2.9e+01&	1.7e-02&     (0.10, 0.40)&  7.2e+01&     6.8e-02&	1.8e+01&     7.0e-01 \\	
(1.2,	1.6)	&(-1.5,	-1.0)	&3.7e+01&	3.4e-02&     (0.10, 0.13)&  1.2e+03&     1.6e-01&	3.8e+02&     7.4e-01 \\	
(1.6,	2.0)	&(-2.5,	-2.0)	&6.1e+01&	7.9e-03&     (0.40, 1.0)&  -1.3e+02&     -5.5e-02&	1.0e+02&     8.1e-01 \\	
(1.6,	2.0)	&(-2.0,	-1.5)	&7.0e+01&	1.9e-02&     (0.13, 1.0)&  9.3e+00&     1.0e-01&	8.4e+00&     8.0e-01 \\	
(1.6,	2.0)	&(-1.5,	-1.0)	&7.5e+01&	4.9e-02&     (0.10, 0.32)&  8.7e+00&     1.8e-01&	4.5e+00&     7.6e-01 \\	
(2.0,	2.4)	&(-2.0,	-1.5)	&1.6e+02&	2.2e-02&     (0.32, 1.0)&  6.1e+00&     1.5e-01&	1.2e+01&     8.7e-01 \\	
(2.0,	2.4)	&(-1.5,	-1.0)	&1.7e+02&	5.8e-02&     (0.10, 0.79)&  1.1e+00&     2.0e-01&	1.5e+00&     7.8e-01 \\	
(2.0,	2.4)	&(-1.0,	-0.5)	&1.9e+02&	1.4e-01&     (0.10, 0.25)&  3.4e+00&     1.8e-01&	2.1e+00&     5.6e-01 \\	
(2.4,	2.8)	&(-2.0,	-1.5)	&2.8e+02&	2.9e-02&     (0.80, 1.0)&  1.3e+02&     1.9e-01&	3.5e+02&     8.9e-01 \\	
(2.4,	2.8)	&(-1.5,	-1.0)	&4.1e+02&	6.3e-02&     (0.25, 1.0)&  5.7e-01&     2.1e-01&	1.7e+00&     7.9e-01 \\	
(2.4,	2.8)	&(-1.0,	-0.5)	&4.4e+02&	1.7e-01&     (0.10, 0.63)&  3.2e-01&     1.6e-01&	4.5e-01&     4.8e-01 \\	
(2.4,	2.8)	&(-0.5,	0.0)	&5.1e+02&	3.8e-01&     (0.10, 0.20)&  4.1e+00&     6.0e-02&	2.7e+00&     1.6e-01 \\	
(2.8,	3.2)	&(-1.5,	-1.0)	&7.6e+02&	8.5e-02&     (0.64, 1.0)&  3.1e+00&     2.2e-01&	1.2e+01&     7.2e-01 \\	
(2.8,	3.2)	&(-1.0,	-0.5)	&1.1e+03&	1.8e-01&     (0.20, 1.0)&  1.2e-01&     1.5e-01&	4.2e-01&     4.5e-01 \\	
(2.8,	3.2)	&(-0.5,	0.0)	&1.1e+03&	4.2e-01&     (0.10, 0.49)&  3.7e-01&     4.7e-02&	5.7e-01&     1.3e-01 \\	
(3.2,	3.6)	&(-1.0,	-0.5)	&2.1e+03&	2.5e-01&     (0.51, 1.0)&  3.2e-01&     1.1e-01&	1.4e+00&     3.0e-01 \\	
(3.2,	3.6)	&(-0.5,	0.0)	&2.7e+03&	4.2e-01&     (0.20, 1.0)&  9.5e-02&     4.3e-02&	3.7e-01&     1.2e-01 \\	
(3.6,	4.0)	&(-0.5,	0.0)	&5.1e+03&	5.7e-01&     (0.59, 1.0)&  4.6e-01&     1.3e-02&	2.2e+00&     3.4e-02 \\  \hline	
\end{tabular}
\caption{
Anticipated sensitivities of $g_1^{\gamma Z}$ and $g_5^{\gamma Z}$ functions for individual bins in the ($Q^2$, $x$) plane.
The projections are for the 10 GeV unpolarized electron beam on the 250 GeV longitudinally polarized proton beam.
The first two columns define the ($\log_{10}(Q^2)$, $\log_{10}(x)$) two dimensional bins.
The $\langle Q^2 \rangle$ and $\langle x \rangle$ are the $f^2(Q^2)$ weighted (as discussed in Sec. \ref{simulation}) mean values in each bin.
The $y$ coverage for the bin is also tabulated.
The $\langle g_1^{\gamma Z} \rangle$ and $\langle g_5^{\gamma Z} \rangle$ are the predicted mean values for the structure functions.
The $\frac{\sigma_{g_1^{\gamma Z}}}{g_1^{\gamma Z}}$ and  $\frac{\sigma_{g_5^{\gamma Z}}}{g_5^{\gamma Z}}$ are the projected relative uncertainties.
The cuts mentioned in Sec. \ref{simulation} are applied to the data.
}
\label{table_g_projection_10_250}
\end{center}
\end{table*}

\begin{table*}
\begin{center}
\begin{tabular}{|c|c|c|c|c|c|c|c|c|}\hline
     $\log_{10}(Q^2)$ bin  &      $\log_{10}(x)$ bin &    $\langle Q^2 \rangle$ (GeV$^2$) &   $\langle x \rangle$   &   $y$ coverage   & $\frac{\sigma_{g_1^{\gamma Z}}}{g_1^{\gamma Z}}$ & $\langle g_1^{\gamma Z} \rangle$    &  $\frac{\sigma_{g_5^{\gamma Z}}}{g_5^{\gamma Z}}$   & $\langle g_5^{\gamma Z} \rangle$        \\      \hline  \hline
(0.0,	0.4)	&(-4.0,	-3.5)	&1.2e+00&	2.8e-04&     (0.53, 0.80)&  -5.7e+02&     -6.8e+00&	5.4e+04&     8.0e-01 \\	
(0.0,	0.4)	&(-3.5,	-3.0)	&1.7e+00&	6.4e-04&     (0.17, 0.80)&  -4.5e+01&     -3.4e+00&	3.0e+03&     4.0e-01 \\	
(0.0,	0.4)	&(-3.0,	-2.5)	&1.9e+00&	1.8e-03&     (0.10, 0.42)&  -9.0e+01&     -1.3e+00&	1.9e+03&     1.6e-01 \\	
(0.0,	0.4)	&(-2.5,	-2.0)	&2.3e+00&	3.5e-03&     (0.10, 0.13)&  -6.5e+03&     -5.5e-01&	3.0e+04&     1.8e-01 \\	
(0.4,	0.8)	&(-3.5,	-3.0)	&3.3e+00&	8.3e-04&     (0.42, 0.81)&  -1.8e+02&     -2.4e+00&	9.7e+03&     4.7e-01 \\	
(0.4,	0.8)	&(-3.0,	-2.5)	&4.4e+00&	2.0e-03&     (0.13, 0.81)&  -3.3e+01&     -1.1e+00&	6.5e+02&     3.8e-01 \\	
(0.4,	0.8)	&(-2.5,	-2.0)	&4.8e+00&	5.2e-03&     (0.10, 0.33)&  -1.5e+02&     -3.3e-01&	3.0e+02&     3.6e-01 \\	
(0.4,	0.8)	&(-2.0,	-1.5)	&6.2e+00&	1.0e-02&     (0.10, 0.11)&  -4.3e+05&     -6.2e-02&	8.5e+04&     4.5e-01 \\	
(0.8,	1.2)	&(-3.0,	-2.5)	&9.2e+00&	2.5e-03&     (0.33, 0.82)&  -7.0e+01&     -7.5e-01&	1.0e+03&     5.2e-01 \\	
(0.8,	1.2)	&(-2.5,	-2.0)	&1.1e+01&	6.0e-03&     (0.11, 0.82)&  -3.6e+01&     -2.4e-01&	9.2e+01&     5.3e-01 \\	
(0.8,	1.2)	&(-2.0,	-1.5)	&1.2e+01&	1.5e-02&     (0.10, 0.26)&  1.0e+03&     2.1e-02&	7.2e+01&     5.8e-01 \\	
(1.2,	1.6)	&(-2.5,	-2.0)	&2.5e+01&	7.2e-03&     (0.26, 0.84)&  -6.0e+01&     -1.2e-01&	1.1e+02&     6.7e-01 \\	
(1.2,	1.6)	&(-2.0,	-1.5)	&2.8e+01&	1.8e-02&     (0.10, 0.66)&  2.9e+01&     7.6e-02&	1.4e+01&     7.0e-01 \\	
(1.2,	1.6)	&(-1.5,	-1.0)	&3.2e+01&	4.1e-02&     (0.10, 0.21)&  5.7e+01&     1.6e-01&	2.3e+01&     7.2e-01 \\	
(1.6,	2.0)	&(-2.5,	-2.0)	&4.4e+01&	9.3e-03&     (0.67, 0.86)&  -6.9e+03&     -1.6e-02&	1.8e+03&     7.6e-01 \\	
(1.6,	2.0)	&(-2.0,	-1.5)	&6.5e+01&	2.1e-02&     (0.21, 0.91)&  9.3e+00&     1.2e-01&	1.3e+01&     7.9e-01 \\	
(1.6,	2.0)	&(-1.5,	-1.0)	&7.1e+01&	5.7e-02&     (0.10, 0.53)&  3.0e+00&     1.9e-01&	2.5e+00&     7.4e-01 \\	
(1.6,	2.0)	&(-1.0,	-0.5)	&8.6e+01&	1.2e-01&     (0.10, 0.17)&  2.6e+01&     1.9e-01&	1.3e+01&     6.0e-01 \\	
(2.0,	2.4)	&(-2.0,	-1.5)	&1.3e+02&	2.7e-02&     (0.53, 1.0)&  1.7e+01&     1.7e-01&	4.3e+01&     8.4e-01 \\	
(2.0,	2.4)	&(-1.5,	-1.0)	&1.7e+02&	5.9e-02&     (0.17, 1.0)&  7.5e-01&     2.0e-01&	1.8e+00&     7.7e-01 \\	
(2.0,	2.4)	&(-1.0,	-0.5)	&1.8e+02&	1.6e-01&     (0.10, 0.42)&  9.4e-01&     1.7e-01&	8.4e-01&     5.1e-01 \\	
(2.0,	2.4)	&(-0.5,	0.0)	&2.3e+02&	3.5e-01&     (0.10, 0.13)&  4.7e+01&     7.2e-02&	2.6e+01&     2.0e-01 \\	
(2.4,	2.8)	&(-1.5,	-1.0)	&3.6e+02&	7.6e-02&     (0.42, 1.0)&  1.1e+00&     2.1e-01&	4.0e+00&     7.4e-01 \\	
(2.4,	2.8)	&(-1.0,	-0.5)	&4.4e+02&	1.8e-01&     (0.13, 1.0)&  1.8e-01&     1.6e-01&	4.6e-01&     4.8e-01 \\	
(2.4,	2.8)	&(-0.5,	0.0)	&4.5e+02&	4.1e-01&     (0.10, 0.33)&  9.6e-01&     5.3e-02&	9.5e-01&     1.4e-01 \\	
(2.8,	3.2)	&(-1.0,	-0.5)	&9.6e+02&	2.2e-01&     (0.34, 1.0)&  2.1e-01&     1.3e-01&	8.7e-01&     3.8e-01 \\	
(2.8,	3.2)	&(-0.5,	0.0)	&1.1e+03&	4.2e-01&     (0.13, 0.83)&  1.8e-01&     4.6e-02&	4.9e-01&     1.3e-01 \\	
(3.2,	3.6)	&(-1.0,	-0.5)	&1.7e+03&	3.0e-01&     (0.86, 1.0)&  1.5e+01&     8.4e-02&	7.2e+01&     2.3e-01 \\	
(3.2,	3.6)	&(-0.5,	0.0)	&2.2e+03&	4.7e-01&     (0.34, 1.0)&  2.1e-01&     3.3e-02&	9.6e-01&     8.8e-02 \\ \hline	
\end{tabular}
\caption{
Anticipated sensitivities of $g_1^{\gamma Z}$ and $g_5^{\gamma Z}$ functions for individual bins in the ($Q^2$, $x$) plane.
The projections are for the 15 GeV unpolarized electron beam on the 100 GeV longitudinally polarized proton beam.
The first two columns define the ($\log_{10}(Q^2)$, $\log_{10}(x)$) two dimensional bins.
The $\langle Q^2 \rangle$ and $\langle x \rangle$ are the $f^2(Q^2)$ weighted (as discussed in Sec. \ref{simulation}) mean values in each bin.
The $y$ coverage for the bin is also tabulated.
The $\langle g_1^{\gamma Z} \rangle$ and $\langle g_5^{\gamma Z} \rangle$ are the predicted mean values for the structure functions.
The $\frac{\sigma_{g_1^{\gamma Z}}}{g_1^{\gamma Z}}$ and  $\frac{\sigma_{g_5^{\gamma Z}}}{g_5^{\gamma Z}}$ are the projected relative uncertainties.
The cuts mentioned in Sec. \ref{simulation} are applied to the data.
}
\label{table_g_projection_15_100}
\end{center}
\end{table*}

\begin{table*}
\begin{center}
\begin{tabular}{|c|c|c|c|c|c|c|c|c|}\hline
     $\log_{10}(Q^2)$ bin  &       $\log_{10}(x)$ bin &  $\langle Q^2 \rangle$ (GeV$^2$) &    $\langle x \rangle$     &  $y$ coverage  &  $\frac{\sigma_{g_1^{\gamma Z}}}{g_1^{\gamma Z}}$ & $\langle g_1^{\gamma Z} \rangle$    &  $\frac{\sigma_{g_5^{\gamma Z}}}{g_5^{\gamma Z}}$   & $\langle g_5^{\gamma Z} \rangle$     \\      \hline  \hline
(0.0,	0.4)	&(-4.5,	-4.0)	&1.1e+00&	9.4e-05&     (0.67, 0.80)&  -4.9e+03&     -1.7e+01&	4.0e+05&     2.5e+00 \\	
(0.0,	0.4)	&(-4.0,	-3.5)	&1.6e+00&	2.1e-04&     (0.21, 0.80)&  -6.8e+01&     -8.9e+00&	4.2e+03&     1.3e+00 \\	
(0.0,	0.4)	&(-3.5,	-3.0)	&1.8e+00&	5.8e-04&     (0.10, 0.53)&  -7.5e+01&     -3.7e+00&	2.0e+03&     4.8e-01 \\	
(0.0,	0.4)	&(-3.0,	-2.5)	&2.2e+00&	1.2e-03&     (0.10, 0.17)&  -1.5e+03&     -1.7e+00&	1.6e+04&     2.5e-01 \\	
(0.4,	0.8)	&(-4.0,	-3.5)	&2.9e+00&	2.8e-04&     (0.53, 0.80)&  -5.6e+02&     -6.5e+00&	3.8e+04&     1.1e+00 \\	
(0.4,	0.8)	&(-3.5,	-3.0)	&4.3e+00&	6.5e-04&     (0.17, 0.81)&  -4.0e+01&     -3.2e+00&	1.5e+03&     6.7e-01 \\	
(0.4,	0.8)	&(-3.0,	-2.5)	&4.7e+00&	1.8e-03&     (0.10, 0.42)&  -8.0e+01&     -1.2e+00&	6.5e+02&     4.1e-01 \\	
(0.4,	0.8)	&(-2.5,	-2.0)	&5.8e+00&	3.5e-03&     (0.10, 0.13)&  -5.6e+03&     -4.9e-01&	1.0e+04&     4.0e-01 \\	
(0.8,	1.2)	&(-3.5,	-3.0)	&8.2e+00&	8.3e-04&     (0.42, 0.81)&  -1.4e+02&     -2.4e+00&	4.9e+03&     7.3e-01 \\	
(0.8,	1.2)	&(-3.0,	-2.5)	&1.1e+01&	1.9e-03&     (0.13, 0.82)&  -2.6e+01&     -1.1e+00&	3.2e+02&     6.0e-01 \\	
(0.8,	1.2)	&(-2.5,	-2.0)	&1.2e+01&	5.1e-03&     (0.10, 0.33)&  -1.3e+02&     -2.9e-01&	1.5e+02&     5.5e-01 \\	
(0.8,	1.2)	&(-2.0,	-1.5)	&1.6e+01&	1.0e-02&     (0.10, 0.11)&  -5.9e+05&     -2.6e-02&	3.6e+04&     6.1e-01 \\	
(1.2,	1.6)	&(-3.0,	-2.5)	&2.3e+01&	2.5e-03&     (0.33, 0.84)&  -5.1e+01&     -7.2e-01&	5.2e+02&     7.3e-01 \\	
(1.2,	1.6)	&(-2.5,	-2.0)	&2.8e+01&	5.9e-03&     (0.11, 0.84)&  -3.0e+01&     -2.1e-01&	5.1e+01&     7.0e-01 \\	
(1.2,	1.6)	&(-2.0,	-1.5)	&3.1e+01&	1.4e-02&     (0.10, 0.27)&  2.9e+02&     5.1e-02&	4.1e+01&     7.1e-01 \\	
(1.6,	2.0)	&(-3.0,	-2.5)	&4.0e+01&	3.2e-03&     (0.84, 0.84)&  -2.0e+06&     -4.8e-01&	1.5e+07&     8.1e-01 \\	
(1.6,	2.0)	&(-2.5,	-2.0)	&6.3e+01&	7.0e-03&     (0.27, 0.91)&  -4.7e+01&     -1.0e-01&	5.8e+01&     8.2e-01 \\	
(1.6,	2.0)	&(-2.0,	-1.5)	&7.0e+01&	1.8e-02&     (0.10, 0.67)&  1.4e+01&     1.0e-01&	8.0e+00&     8.0e-01 \\	
(1.6,	2.0)	&(-1.5,	-1.0)	&8.1e+01&	4.1e-02&     (0.10, 0.21)&  3.2e+01&     1.8e-01&	1.3e+01&     7.8e-01 \\	
(2.0,	2.4)	&(-2.5,	-2.0)	&1.1e+02&	9.0e-03&     (0.67, 0.96)&  4.6e+03&     8.6e-03&	5.6e+02&     8.9e-01 \\	
(2.0,	2.4)	&(-2.0,	-1.5)	&1.7e+02&	1.9e-02&     (0.21, 1.0)&  4.6e+00&     1.3e-01&	7.3e+00&     8.9e-01 \\	
(2.0,	2.4)	&(-1.5,	-1.0)	&1.8e+02&	5.6e-02&     (0.10, 0.53)&  1.8e+00&     2.0e-01&	1.6e+00&     7.8e-01 \\	
(2.0,	2.4)	&(-1.0,	-0.5)	&2.1e+02&	1.2e-01&     (0.10, 0.17)&  1.6e+01&     1.9e-01&	8.1e+00&     6.0e-01 \\	
(2.4,	2.8)	&(-2.0,	-1.5)	&3.3e+02&	2.6e-02&     (0.53, 1.0)&  9.7e+00&     1.8e-01&	2.5e+01&     9.2e-01 \\	
(2.4,	2.8)	&(-1.5,	-1.0)	&4.3e+02&	5.8e-02&     (0.17, 1.0)&  4.5e-01&     2.1e-01&	1.1e+00&     8.1e-01 \\	
(2.4,	2.8)	&(-1.0,	-0.5)	&4.5e+02&	1.6e-01&     (0.10, 0.42)&  6.4e-01&     1.7e-01&	5.7e-01&     5.0e-01 \\	
(2.4,	2.8)	&(-0.5,	0.0)	&5.7e+02&	3.5e-01&     (0.10, 0.13)&  3.5e+01&     7.1e-02&	1.9e+01&     2.0e-01 \\	
(2.8,	3.2)	&(-1.5,	-1.0)	&9.1e+02&	7.5e-02&     (0.42, 1.0)&  7.4e-01&     2.2e-01&	2.7e+00&     7.6e-01 \\	
(2.8,	3.2)	&(-1.0,	-0.5)	&1.1e+03&	1.7e-01&     (0.14, 1.0)&  1.2e-01&     1.6e-01&	3.0e-01&     4.8e-01 \\	
(2.8,	3.2)	&(-0.5,	0.0)	&1.1e+03&	4.1e-01&     (0.10, 3.3)&  6.6e-01&     5.1e-02&	6.6e-01&     1.4e-01 \\	
(3.2,	3.6)	&(-1.0,	-0.5)	&2.4e+03&	2.2e-01&     (0.34, 1.0)&  1.3e-01&     1.3e-01&	5.3e-01&     3.7e-01 \\	
(3.2,	3.6)	&(-0.5,	0.0)	&2.8e+03&	4.2e-01&     (0.15, 0.83)&  1.1e-01&     4.4e-02&	3.1e-01&     1.2e-01 \\	
(3.6,	4.0)	&(-1.0,	-0.5)	&4.3e+03&	2.9e-01&     (0.90, 1.0)&  2.4e+01&     8.1e-02&	1.2e+02&     2.2e-01 \\	
(3.6,	4.0)	&(-0.5,	0.0)	&5.5e+03&	4.5e-01&     (0.39, 1.0)&  1.4e-01&     3.4e-02&	6.6e-01&     9.1e-02 \\ \hline	
\end{tabular}
\caption{
Anticipated sensitivities of $g_1^{\gamma Z}$ and $g_5^{\gamma Z}$ functions for individual bins in the ($Q^2$, $x$) plane.
The projections are for the 15 GeV unpolarized electron beam on the 250 GeV longitudinally polarized proton beam.
The first two columns define the ($\log_{10}(Q^2)$, $\log_{10}(x)$) two dimensional bins.
The $\langle Q^2 \rangle$ and $\langle x \rangle$ are the $f^2(Q^2)$ weighted (as discussed in Sec. \ref{simulation}) mean values in each bin.
The $y$ coverage for the bin is also tabulated.
The $\langle g_1^{\gamma Z} \rangle$ and $\langle g_5^{\gamma Z} \rangle$ are the predicted mean values for the structure functions.
The $\frac{\sigma_{g_1^{\gamma Z}}}{g_1^{\gamma Z}}$ and  $\frac{\sigma_{g_5^{\gamma Z}}}{g_5^{\gamma Z}}$ are the projected relative uncertainties.
The cuts mentioned in Sec. \ref{simulation} are applied to the data.
}
\label{table_g_projection_15_250}
\end{center}
\end{table*}

\section*{Acknowledgements}
The authors are grateful to Hubert Spiesberger for useful discussions and advice on the DJANGOH generator, and Jens Erler for information on the running of the 
weak mixing angle in the $\overline{\rm{MS}}$ scheme. This work was supported by Department of Energy (DOE) under contract numbers DE-SC00013321 and DE-FG-02-05-ER41372 (SBU), and DE-SC0012704 (BNL).



\end{document}